\documentclass[11pt]{article}
\usepackage{amsmath,amssymb,color,epsfig,cite,CJK}
\usepackage{graphicx}
\usepackage{subfigure}
\usepackage{setspace}
\usepackage{braket}

\usepackage{amsfonts}

\newcommand{\hoch}[1]{$\, ^{#1}$}

\makeatletter
\@addtoreset{equation}{section}
\makeatother

\newcommand{\be}{\begin{equation}}
\newcommand{\ee}{\end{equation}}
\newcommand{\bea}{\setlength\arraycolsep{2pt} \begin{eqnarray}}
\newcommand{\eea}{\end{eqnarray}}
\newcommand{\nn}{\nonumber}

\newcommand{\bpm}{\begin{pmatrix}}
\newcommand{\epm}{\end{pmatrix}}

\def\ft#1#2{{\textstyle{\frac{\scriptstyle #1}{\scriptstyle #2} } }}
\def\fft#1#2{{\frac{#1}{#2}}}

\def\0{{\sst{(0)}}}
\def\1{{\sst{(1)}}}
\def\2{{\sst{(2)}}}
\def\3{{\sst{(3)}}}
\def\4{{\sst{(4)}}}
\def\5{{\sst{(5)}}}
\def\6{{\sst{(6)}}}
\def\7{{\sst{(7)}}}
\def\8{{\sst{(8)}}}
\def\sst#1{{\scriptscriptstyle #1}}

\begin{document}


\begin{center}
{\large {\bf Holographic Complexity Bounds}}

\vspace{10pt}
Hai-Shan Liu\hoch{1,2}, H. L\"u\hoch{1*}, Liang Ma\hoch{1} and Wen-Di Tan\hoch{1}

\vspace{10pt}

\hoch{1}{\it Center for Joint Quantum Studies and Department of Physics,\\
School of Science, Tianjin University, Tianjin 300350, China}

\vspace{10pt}

\hoch{2}{\it Institute for Advanced Physics \& Mathematics,\\
Zhejiang University of Technology, Hangzhou 310023, China}

\vspace{40pt}

\underline{ABSTRACT}
\end{center}

We study the action growth rate in the Wheeler-DeWitt (WDW) patch for a variety of $D\ge 4$ black holes in Einstein gravity that are asymptotic to the anti-de Sitter spacetime, with spherical, toric and hyperbolic horizons, corresponding to the topological parameter $k=1,0,-1$ respectively. We find a lower bound inequality $\fft{1}{T} \fft{\partial \dot I _{\rm WDW}}{\partial S}|_{Q,P_{\rm th}}> C$ for $k=0,1$, where $C$ is some order-one numerical constant.  The lowest number in our examples is $C=(D-3)/(D-2)$.  We also find that the quantity $(\dot I_{\rm WDW}-2P_{\rm th}\, \Delta V_{\rm th})$ is greater than, equal to,  or less than zero, for $k=1,0,-1$ respectively. For black holes with two horizons, $\Delta V_{\rm th}=V_{\rm th}^+-V_{\rm th}^-$, {\it i.e.}~the difference between the thermodynamical volumes of the outer and inner horizons.  For black holes with only one horizon, we introduce a new concept of the volume $V_{\rm th}^0$ of the black hole singularity, and define $\Delta V_{\rm th}=V_{\rm th}^+-V_{\rm th}^0$. The volume $V_{\rm th}^0$ vanishes for the Schwarzschild black hole, but in general it can be positive, negative or even divergent. For black holes with single horizon, we find a relation between $\dot I_{\rm WDW}$ and $V_{\rm th}^0$, which implies that the holographic complexity preserves the Lloyd's bound for positive or vanishing $V_{\rm th}^0$, but the bound is violated when $V_{\rm th}^0$ becomes negative.  We also find explicit black hole examples where $V_{\rm th}^0$ and hence $\dot I_{\rm WDW}$ are divergent.

\vfill {\footnotesize  hsliu.zju@gmail.com \ \ \  mrhonglu@gmail.com\ \ \ liangma@tju.edu.cn\ \ \
twdhannah@163.com}

{\footnotesize \hoch{*}Corresponding author}

\thispagestyle{empty}

\pagebreak

\tableofcontents
\addtocontents{toc}{\protect\setcounter{tocdepth}{2}}

\newpage

\section{Introduction}

The gauge/gravity duality provides a powerful methodology to study the dynamics of various strongly
coupled condensed matter systems \cite{adscft1,adscft2,adscft3,adscft4}. A standard technique is to consider a certain asymptotically anti-de Sitter (AdS) black holes in $D=d+1$ dimensions and derive the $d$-dimensional boundary properties using the holographic dictionary.  The holographic principle can also be adopted to study quantum information and two conjectures about the quantum computational complexity have been proposed.  One is the ``complexity = volume''(CV) conjecture \cite{cv1,cv2} and the other is the ``complexity = action" (CA) conjecture \cite{ca1,ca2}. These conjectures attracted considerable attentions and many works have been done to investigate the properties of holographic complexity associated with these two conjectures \cite{camyers,cav2,cav3,cav4,cav5, cav7,cav8,cav9,cav10,cav11,cav12,cav13, ptwoh, masgr, frcg, fr, frmas, lovelock,kim,feng} and to further generalize these two conjectures\cite{cav1,cav6,jiang,cvt1,sub0}.

Holographic technique appears to be particularly useful to give some universal properties of the boundary field theory. In this regard, the Schwarzschild-AdS black hole plays a unique r\^ole since the results are independent of any matter contents, thus its holographic properties could be universal or provide bounds for more general systems. According to the CA conjecture, the late-time growth rate of the holographic complexity is the growth rate of the on-shell action of the dual AdS black hole in the Wheeler-DeWitt (WDW) patch.  For the Schwarzschild-AdS black hole of mass $M$, it was found to be \cite{ca2}
\be
\dot I_{\rm WDW} = 2M\,.
\ee
This result is consistent with Lloyd's bound on the growth rate of the quantum complexity, namely
\cite{lloyd}
\be
\dot {\cal C} \le {2 E} \,,
\ee
where $E$ is the energy of the quantum system.  The Lloyd's bound is saturated by the Schwarzschild-AdS black hole. For both neutral and charged black holes, an upper bound for the complexity growth was proposed\cite{ca1,ca2}, and further exploration of these bounds were carried out in \cite{cav1,cai1,cai2}. The success tempted one to suggest that the action growth rate is always equal to twice the mass for neutral black holes \cite{ca1,ca2,cai1,cai2}, at least for black holes in Einstein gravity satisfying the null energy condition. The strong energy condition outside the horizon of an eternal neutral black hole (with Schwarzschild interior) can ensure the Lloyd's bound \cite{Yang:2016awy}.  This however is a rather restrictive class of solutions. Indeed it was recently found that the Lloyd's bound could be violated for some black holes in scalar-tensor theory \cite{cav6,Mahapatra:2018gig}; however, there was no underlying explanation as why these black holes are different from the others that do satisfy the Lloyd's bound.

In the CV conjecture, there was an ambiguity of the necessary multiplication factor needed to equate complexity to certain volume, which have different dimensions \cite{cv1,cv2}.  With the proposal \cite{Kastor:2009wy,Cvetic:2010jb} of thermodynamical pressure $P_{\rm th}$ and volume $V_{\rm th}$ in AdS black holes, it is natural to consider that the product $P_{\rm th} V_{\rm th}$, which has the same dimension as energy, could be a candidate for holographic complexity \cite{cav1,cav6}.  It was demonstrated that for general two-horizon black holes, the action growth rate in the WDW patch can always be expressed universally as \cite{ptwoh}
\be
\dot I_{\rm WDW} = {\cal H}_+ - {\cal H}_-\,,\label{genformula}
\ee
namely the difference between the enthalpy associated with the inner and outer horizons. This result implies that $\dot I_{\rm WDW}$ is expressed solely in terms of black hole thermodynamical quantities and hence it allows us to relate CA and the new CV conjectures by making use of the Smarr relations of the thermodynamical variables.

For two-horizon black holes, it is natural to consider $2P_{\rm th}\, \Delta V_{\rm th}$, with $\Delta V_{\rm th}= V_{\rm th}^+ - V_{\rm th}^-$ where $V_{\rm th}^\pm$ are the thermodynamical volumes on the outer and inner horizons. (The proposal in \cite{cav6} does not include the factor 2.) The situation becomes tricky when there is only one horizon, and naively one might simply propose $2P_{\rm th} V_{\rm th}^+$ as the holographic complexity. We find that it is necessary and useful to introduce the concept of the volume of singularity $V_{\rm th}^0$, and define $\Delta V_{\rm th}= V_{\rm th}^+ - V_{\rm th}^0$.  For the Schwarzschild-AdS black hole, $V_{\rm th}^0$ is simply zero. However, for more general black holes $V_{\rm th}^0$ does not have to  vanish. It turns out that both positive and negative values of $V_{\rm th}^0$ can emerge from various black holes. We find that black holes with $V_{\rm th}^0\ge 0$ satisfy the Lloyd's bound and those with $V_{\rm th}^0< 0$ can violate the bound.  Intriguingly we also find explicit examples of black holes with divergent $V_{\rm th}^0$ and $\dot I_{\rm WDW}$.

Since both the CA and our new CV conjectures express the holographic complexity in terms of the black hole thermodynamical variables, we can also explore the differential relation such as the first law of black hole thermodynamics.  For the Schwarzschild-AdS, we have
\be
\fft{1}{T} \fft{\partial \dot I_{\rm WDW} }{\partial S} =\fft{2}{T} \fft{\partial M}{\partial S}= 2\,.
\ee
This equation suggests that complexity growth rate is a monotonic function of the entropy and furthermore the rate of increase is proportional to the temperature. In this paper we would like to explore the possibility whether this is universal or provides a bound for the general ${\cal C}$-$S$ relation.

The paper is organized as follows. In section \ref{sec:EMgravity}, we review the holographic complexity in both Schwarzschild and Reissner-Nordstr\"om (RN) AdS  black holes and obtain both the algebraic and differential inequalities.  We then present the general statements of the bounds in section \ref{sec:bounds}. We prove the saturation of some bounds in the case of $k=0$.  In sections \ref{sec:emd} and \ref{sec:EBI}, we study AdS black holes Einstein gravities with a variety of matter all satisfying the null energy condition, and verify the bounds with explicit examples.  In section \ref{sec:negvol}, we focus on the study of volume of singularity and provide explicit examples of black holes with negative or even divergent singularity volumes. We study the consequent effects on the holographic complexity. The negative volume leads to the violation of the Lloyd's bound and the divergent one gives rise to divergent action growth rate in the WDW patch.  We conclude the paper in section \ref{sec:con}.

\section{Complexity bounds from Einstein-Maxwell theory}
\label{sec:EMgravity}

In this section, we consider Einstein-Maxwell gravity coupled to a negative cosmological constant $\Lambda$ in general $D$ dimensions, with the Lagrangian
\be
{\cal L}=\sqrt{-g} (R - \ft14 F^2 - 2\Lambda)\,,
\ee
where $F=dA$ is the Maxwell field strength and $A$ is the gauge potential.  For negative $\Lambda$, the theory admits both the (neutral) Schwarzschild-AdS black hole and (charged) RN-AdS black hole.  The holographic complexity associated with these black holes based on the CA conjecture was well known.  Based on these results and further general conditions, we propose general bounds on the holographic complexity.

\subsection{Schwarzschild-AdS black hole}
\label{sec:schwarz}

The Schwarzschild-AdS metric in general $D$ dimensions is
\be
ds^2_D = - f dt^2 + \fft{dr^2}{f} + r^2 d\Omega_{D-2,k}^2\,,\qquad
f=g^2 r^2 + k -\fft{\mu}{r^{D-3}}\,,
\ee
where $d\Omega_{D-2,k}^2$ is the metric for maximally-symmetric space with $R_{ij}=(D-3)k g_{ij}$, and $k=1,0,-1$, corresponding to unit sphere $S^{D-2}$, torus $T^{D-2}$ and hyperbolic $H^{D-2}$.  The metric is asymptotic to AdS spacetimes with radius $\ell=1/g$, where we parameterize the cosmological constant as
\be
\Lambda=-\ft12 (D-1)(D-2) g^2\,.
\ee
Note that in this paper there should be no confusion between the parameter $g$ and the metric determinant.
For suitable $\mu$, the metric has an event horizon $r_+>0$ with $f(r_+)=0$ and the metric describes a black hole, satisfying the first law of black hole thermodynamics
\be
dM=TdS + V_{\rm th} dP_{\rm th}\,,\label{fl0}
\ee
where the thermodynamical variables are
\bea
M &=& \fft{(D-2)\Omega_{D-2,k}}{16\pi} \mu\,,\qquad T=\fft{f'(r_+)}{4\pi}\,,\qquad S=\ft14 \Omega_{D-2,k} r_+^{D-2}\,,\nn\\
P_{\rm th}&=& -\fft{\Lambda_0}{8\pi}=\fft{(D-1)(D-2)}{16\pi} g^2\,,\qquad
V_{\rm th}=\fft{\Omega_{D-2,k}}{D-1}r_+^{D-1}\,.
\eea
Here $\Omega_{D-2}$ denotes the volume of the unit $S^{D-2}$, (corresponding to $k=1$,) given by
\be
\Omega_{D-2} = \fft{(2\pi)^{(D-1)/2}}{\Gamma[\ft12 (D-1)]}\,,
\ee
with $\Omega_2=4\pi$, $\Omega_{3}=2\pi^2$, $\Omega_4=\ft83 \pi^2$ and $\Omega_5=\pi^3$.  For $k=0$ or $-1$, the metric $d\Omega_{D-2,k}^2$ is not compact, but for simplicity we still assign the same value of $\Omega_{D-2}$ as if it were a unit sphere, and therefore we shall drop the label $k$ and use simply $\Omega_{D-2}$ for the volume of $d\Omega_{D-2,k}^2$ for all topologies. It should be kept in mind that the extensive quantities such as mass, charge and entropy should be understood as densities when $k=0$ or $-1$.

We also followed \cite{Kastor:2009wy,Cvetic:2010jb} and treated the negative cosmological constant $\Lambda$ as positive pressure $P_{\rm th} = \Lambda/(8\pi)$.  Its conjugate is the thermodynamical volume $V_{\rm th}$, which in this case is simply the
Euclidean volume of a spherical ball of radius $r_+$. The first law (\ref{fl0}) indicates that the mass of the AdS black hole should be really viewed as enthalpy ${\cal H}$ rather than the internal energy in the black hole thermodynamical system. As we shall see throughout the paper, black hole volume is a key element for understanding the holographic complexity.

For the Schwarzschild-AdS black holes, we find the following algebraic identity and the consequent inequality
\be
2M - 2 P_{\rm th} V_{\rm th} = k \fft{(D-2)\Omega_{D-2}}{8\pi} r_+^{D-3} \quad \rightarrow\quad
\left\{
  \begin{array}{ll}
    >0, &\qquad k=1\,; \\
    =0, &\qquad k=0\,; \\
    <0, &\qquad k=-1\,.
  \end{array}
\right.
\ee
Furthermore, we have the differential inequality
\be
\fft{\partial M}{\partial S}\Big|_{P_{\rm th}}=T>0\,.
\ee
Based on the CA conjecture, the holographic later time complexity is given by the on-shell action in the WDW patch. For the Schwarzschild-AdS black hole, the action growth rate is simply twice the mass, namely \cite{ca1,ca2}
\be
\dot I_{\rm WDW}=2M.
\ee
It follows that the above thermodynamical inequalities become those of later time growth rate of the complexity.  It is also of interest to note that
\be
\fft{\partial (2P_{\rm th} V_{\rm th})}{\partial S}\Big|_{P_{\rm th}} = \frac{(D-1) g^2 r_+}{2 \pi }>0\,.
\ee
Although this quantity is always positive, it can approach zero at high temperature for small black holes with $r_+\rightarrow 0$.

\subsection{RN-AdS black hole}
\label{sec:RN}

The information we can learn from the Schwarzschild-AdS black hole is limited. We thus progress to study the thermodynamic inequalities in RN-AdS black hole to search for some general patterns. The charged AdS black hole solution is
\bea
ds_{D}^2 &=& - f dt^2 + \fft{dr^2}{f} + r^2 d\Omega_{D-2,k}^2\,,
\qquad A=\fft{q}{(D-3) r^{D-3}} dt\,,\nn\\
f &=& g^2 r^2 + k - \fft{\mu}{r^{D-3}} + \fft{ q^2}{2(D-2)(D-3)r^{2(D-3)}}\,,
\eea
The solution contains two integration constants $(\mu, q)$, parameterizing the conserved quantities, mass and electric charge
\be
M=\fft{(D-2)\Omega_{D-2}}{16\pi} \mu\,,\qquad
Q=\fft{\Omega_{D-2}}{16\pi} q\,.
\ee
For appropriate choices of the parameter, the solution describes a black hole with two horizons, the inner $r_->0$ and outer $r_+\ge r_-$, with $f(r_\pm)=0$. In this paper, we use a real $r_0>0$ with $f(r_0)=0$ to denote a generic horizon which can be either the inner or the outer.  The rest of the thermodynamical quantities are
\bea
T&=&\fft{f'(r_0)}{4\pi}\,,\qquad S=\ft14 \Omega_{D-2} r_0^{D-2}\,,\qquad
\Phi=\fft{q}{(D-3)r_0^{D-3}}\nn\\
P_{\rm th}&=&\fft{(D-1)(D-2)}{16\pi} g^2\,,\qquad V_{\rm th}=\fft{\Omega_{D-2}}{D-1}r_0^{D-1}\,.
\eea
The first law of black hole thermodynamics is
\be
dM=TdS + \Phi dQ + V_{\rm th} dP_{\rm th}\,.
\ee
The first law holds for thermodynamical quantities evaluated in both inner and outer horizons.

The action growth rate of the RN-AdS black hole in the WDW patch is \cite{ca2}
\be
\dot I_{\rm WDW} = \Phi Q\Big|_{r_+}^{r_-}\,.
\ee
First we note that
\be
\dot {I}_{\rm WDW} - 2M=- \fft{(D-2) \Omega}{4\pi}
\Big(k + g^2 r_-^2\,\fft{ (\eta^{D-1} + \eta^{D-3} -2)}{2(\eta^{D-3}-1)}\Big) r_-^{D-3}\,,
\ee
where $\eta=r_+/r_-\ge 1$.  This implies
\be
\dot {I}_{\rm WDW}\le 2M=\dot I_{\rm WDW}^{\rm Schwarzschild}\,, \qquad\hbox{for}\qquad k=0,1.\label{ineq2}
\ee
However, the above inequality no longer holds when $k=-1$. A concrete example of violation for $k=-1$ can be provided with
\be
g^2=1,\quad k=-1,\quad D=4, \quad M=\ft{33}{250},\quad Q=\sqrt{\ft{363}{40000}},\quad r_-=\ft1{10}\,,\quad r_+=\ft{11}{10}\,,
\ee
for which $\dot {I}_{\rm WDW} - 2M=\ft{33}{500}>0$. For $k=0,1$, we can also provide a lower bound for $\dot I_{\rm WDW}$. We note that
\be
\dot {I}_{\rm WDW} -2 P_{\rm th}\, \Delta V_{\rm th} = \fft{(D-2)\Omega_{D-2}}{8\pi} k\,
(\eta^{D-3} -1) r_-^{D-3}\quad
\left\{
  \begin{array}{ll}
    >0, &\qquad k=1, \\
    =0, &\qquad k=0, \\
    <0, &\qquad k=-1,
  \end{array}
\right.
\ee
where $\Delta V_{\rm th} = V_{\rm th}\Big|^{r_+}_{r_-}$.

Combining the Lloyd's bound and the above relation between the $\dot I_{\rm WDW}$ and $2P_{\rm th}\, \Delta V_{\rm th}$, we have the algebraic inequalities
\bea
k=0,1:&&\qquad  2 P_{\rm th}\, \Delta V_{\rm th}\, \le\, \dot I_{\rm WDW}\,\le\, 2M,\nn\\
k=-1:&&\qquad  0\,\le\, \dot I_{\rm WDW}\,\le\, 2 P_{\rm th}\, \Delta V_{\rm th}\,.
\eea
Note that for extremal black holes, both $\dot I_{\rm WDW}$ and $\Delta V_{\rm th}$ vanish.

We now turn to the differential inequality.  We present the analysis in $D=4$ since the expressions for general $D$ are quite messy. Motivated by the Schwarzschild-AdS example, we consider
\be
\fft{1}{T} \fft{\partial \dot I_{\rm WDW}}{\partial S}\Big|_{Q,P_{\rm th}}-2=\fft{X}{Y}\,,
\ee
where the quantities $(X,Y)$ can be obtained explicitly, given by
\bea
X&=& 2 r_- \Big(g^4 \left(6 r_-^4+11 r_+ r_-^3+15 r_+^2 r_-^2+3 r_+^3 r_-+r_+^4\right)\nn\\
&&\qquad +g^2 k \left(8 r_-^2+5 r_+ r_-+5 r_+^2\right)+2 k^2\Big)\,,\nn\\
Y&=& \left(r_+-r_-\right) \left(g^2 \left(3 r_-^2+2 r_+ r_-+r_+^2\right)+k\right) \left(g^2 \left(r_-^2+2 r_+ r_-+3 r_+^2\right)+k\right).\label{rndotI}
\eea
We thus see that for $k=0$ or 1, the quantity $X/Y\ge 0$, with the inequality saturated by $r_-=0$ or $r_+\rightarrow \infty$, corresponding the Schwarzschild-AdS black hole, or the charged black hole with infinity mass, respectively. Although the above discussion was presented for $D=4$, we have verified that the conclusion holds for general $D\ge 4$ dimensions.

For $k=-1$, the situation becomes complicated and $X/Y$ can be negative.  As a concrete example, we consider $D=4$ with $g=1, q=1, \mu =1$, for which the inner and outer horizons are $(r_-, r_+)=(0.208, 1.277)$.  In other words, the black hole is well defined; however, we find that $X/Y=-0.180$ in this ($k=-1$) case.  It is still worth asking whether there exists a lower bound $(<2)$ for $k=-1$.  We find
\bea
&&\fft{1}{T} \fft{\partial \dot I_{\rm WDW}}{\partial S}\Big|_{Q,P_{\rm th}}\nn\\
&&=\frac{2 \left(r_-+r_+\right) \left(g^2 r_-^2+g^2 r_+^2+g^2 r_- r_+-1\right) \left(3 g^2 r_-^2+3 g^2 r_+^2-1\right)}{\left(r_+-r_-\right) \left(3 g^2 r_-^2+g^2 r_+^2+2 g^2 r_- r_+-1\right) \left(g^2 r_-^2+3 g^2 r_+^2+2 g^2 r_- r_+-1\right)}\nn\\
&&=1 + \frac{2 g^4 r_-^3 r_+ \left(r_-+r_+\right)^3+g^2 M \left(r_-+r_+\right)^2 \left(3 r_-^2+r_+ r_-+r_+^2\right)+M^2 \left(3 r_-+r_+\right)}{\left(r_+-r_-\right) \left(g^2 r_- \left(r_-+r_+\right)^2+M\right) \left(g^2 r_+ \left(r_-+r_+\right)^2+M\right)}\nn\\
&&=\frac{2 Q^2 \left(r_-+r_+\right) \left(g^2 r_- r_+ \left(2 r_-^2-r_+ r_-+2 r_+^2\right)+Q^2\right)}{\left(r_+-r_-\right) \left(g^2 r_-^2 r_+ \left(2 r_-+r_+\right)+Q^2\right) \left(g^2 r_- r_+^2 \left(r_-+2 r_+\right)+Q^2\right)}\,.
\eea
To interpret the results, we first note that for the RN black hole with $k=-1$, even when the mass is negative, the solution can still have two horizons.  If we insist that the mass is positive, {\it i.e.} $M\ge 0$, then the quantity above is positive with the lowest value being 1. However, if we allow that $M$ can be negative, but keeping $Q^2\ge 0$ so that the Maxwell field is non-ghost like, then the quantity must be positive, but can arbitrarily approach zero.  Furthermore, as we shall see later, the differential quantity can become negative with certain black hole examples. We therefore shall focus only on the $k=0,1$ cases in this paper for the differential inequalities.

As was discussed in the introduction, we may also consider treating $2 P_{\rm th}\, \Delta V_{\rm th}$ as the holographic complexity.  We can also examine its dependence on the black hole entropy in the analogous way.
We find
\bea
&&\fft{1}{T}\fft{\partial (2P_{\rm th}\,\Delta V_{\rm th})}{\partial S}\Big|_{Q,P_{\rm th}}\nn\\
&=& \fft{6g^2\left(g^2 \left(r_-+r_+\right) \left(r_-^2+r_+^2\right) \left(r_-^2+r_+ r_-+r_+^2\right)+k \left(r_-^3+r_+^3\right)\right)}{(r_+ - r_-)\left(k+g^2 \left(3 r_-^2+2 r_+ r_-+r_+^2\right)\right)
\left(k+g^2 \left(r_-^2+2 r_+ r_-+3 r_+^2\right)\right)}\,.
\eea
The right-hand side reduces to that of (\ref{rndotI}) when $k=0$.  For $k=1$, all we can say is that the right-hand side must be positive, and it can approach zero indefinitely.

\section{General statements of holographic complexity bounds}
\label{sec:bounds}

Having analysed the Schwarzschild-AdS and RN-AdS black holes in the previous section, we are now in the position to state the general holographic complexity bounds based on the CA or the new CV conjectures .  We propose two types of the bounds, the algebraic and the differential.

\subsection{Algebraic CA-CV inequalities}
\label{sec:algbound}

First we consider the algebraic relations between the CA and the new CV conjectures.  For an asymptotically AdS black hole of mass $M$,  we would like to propose the following inequalities
\bea
2P_{\rm th}\, \Delta V_{\rm th}\ \le\  &\dot I_{\rm WDW}&\  \le\  2M\,,\qquad\qquad\,\,\,\, \hbox{for}\qquad k=0,1,\nn\\
0\ \le\ &\dot I_{\rm WDW}&\ \le\ 2P_{\rm th}\, \Delta V_{\rm th}\,,\qquad \hbox{for} \qquad k=-1\,.\label{ineq1}
\eea
The above combines the Lloyd's bound with the relation between the CA and the new CV conjectures, namely
\be
\dot {\cal C}_A=\dot I_{\rm WDW}\,,\qquad \dot {\cal C}_V=2P_{\rm th}\, \Delta V_{\rm th}\,.
\ee
Note that we have already known that the Lloyd's bound can be violated. We nevertheless include the bound here since we would like to check this bound in all our examples to achieve some understanding of the underlying reason behind the violation.

The definition of $\Delta V_{\rm th}$ depends on whether the black hole has two horizons or only has one. When it has both the inner $r_-$ and outer $r_+$ horizons, we define
\be
\Delta V_{\rm th} \equiv V_{\rm th}(r_+) - V_{\rm th}(r_-)\,.
\ee
In other words, it is the difference of the thermodynamical volume between the outer and inner horizons.
It should be mentioned that recently exact black holes satisfying the dominant energy condition with four horizons were constructed in Einstein-Maxwell gravity with quasi-topological extension \cite{Liu:2019rib}. We shall not consider this type of black holes here.

When the black hole has only one horizon $r_+$ that shields a curvature singularity, say at $r=0$, we have
\be
\Delta V_{\rm th} \equiv V_{\rm th}(r_+) - V_{\rm th}^0\,,\qquad V_{\rm th}^0 \equiv V_{\rm th}(0)\,.
\ee
The inequalities (\ref{ineq1}) associated with $\Delta V_{\rm th}$ is saturated when $k=0$ or for extremal black holes with non-zero $k$.

In this paper, we introduce a new concept: the volume of a singularity. For the Schwarzschild-AdS or RN-AdS black holes, the volume $V_{\rm th}^0$ simply vanishes at the curvature singularity. However, we shall see that there exist black holes with non-vanishing $V_{\rm th}^0$ and the quantity plays an important r\^ole in our CV conjecture and the resulting inequality.  However, in order not to interrupt our main story in this section, we shall elaborate the details of the volume of singularity later, with concrete examples. It should be emphasized that our new CV conjecture and the one proposed in \cite{cav1,cav6} are distinguished by the black holes with a single horizon.

Planar AdS black holes ($k=0$) have enhanced scaling symmetry and we can establish $\dot {\cal C}_A=\dot {\cal C}_V$ abstractly. To be concrete, we consider a class of charged AdS black holes, carrying mass $M$ and multiple electric charges $Q_i$ associated with different $U(1)$ fields $A_i$.  The first law of black hole thermodynamics states
\be
dM=T dS + \Phi_i dQ_i + V_{\rm th} dP_{\rm th}\,.\label{genfl}
\ee
We assume that the theory has no more additional dimensionful parameters. Based on the dimensions of the thermodynamical variables, we have the Smarr relation
\be
(D-3) M=(D-2) T S + (D-3) \Phi_i Q_i -2 P_{\rm th} V_{\rm th}\,.\label{smarr1}
\ee
This relation holds for all the $k=1,0,-1$ cases. (If the theory involves additional dimensionful couplings, the Smarr relation needs to be augmented with those parameters.)  When $k=0$, the planar AdS black hole has an additional scaling symmetry, which implies a new generalized Smarr relation \cite{Liu:2015tqa}
\be
(D-1) M = (D-2) (T S +\Phi_i Q_i)\,.\label{smarr2}
\ee
Thus for $k=0$, we have
\be
2M-\Phi_i Q_i = 2 P_{\rm th} V_{\rm th}\,.\label{smarr3}
\ee

With these preliminaries, we examine the growth rate of the action in the WDW patch.  We adopt methods developed in \cite{ca2,camyers} for computing the on-shell action in the WDW patch.  The bulk action can also be computed by methods developed in \cite{Liu:2017kml} and adopted in \cite{cav6}. Three cases emerge.

\bigskip
\noindent{\bf Case 1: Charged black holes with two horizons.} When a black hole has two horizons, a general formula (\ref{genformula}) were given in (\ref{genformula}).  For charged black holes, it means
\be
\dot I_{\rm WDW} = \Phi_i Q_i \Big|^{r_-}_{r_+}\,.
\ee
It follows from (\ref{smarr3}) that
\be
\dot I_{\rm WDW} = 2P_{\rm th}\, \Delta V_{\rm th}\,,\qquad \hbox{for}\qquad k=0\,.
\ee
In other words, the CA and our new CV conjectures yield the identical holographic complexity.
\bigskip

\noindent{\bf Case 2: Charged black holes with one horizon.} The action growth rate takes the form
\be
\dot I_{\rm WDW}=2M - \Phi_i Q_i + \cdots,
\ee
where the ellipses denote the quantities needed such that $\dot I_{\rm WDW}$ must vanish in the $r_+\rightarrow 0$ limit when the solution ceases to be a black hole. When $k=0$, it follows from (\ref{smarr3}) that we must have
\be
\dot I_{\rm WDW}=2M - \Phi_i Q_i - 2 P_{\rm th} V_{\rm th}^0= 2P_{\rm th}\, \Delta V_{\rm th},\qquad \hbox{for}\qquad
k=0\,.
\ee

\noindent{\bf Case 3: Neutral black holes with one horizon.} These solutions typically involve scalar hair. Following the analogous discussion of case 2, we have
\be
\dot I_{\rm WDW}=2M - 2P_{\rm th} V_{\rm th}^0= 2P_{\rm th}\, \Delta V_{\rm th},\qquad \hbox{for}\qquad
k=0\,.
\ee
Thus we see that for $k=0$, we have in general $\dot I_{\rm WDW}=2P_{\rm th}\, \Delta V_{\rm th}$. It is clear that $V_{\rm th}^0$ is important to be introduced in order to avoid the discontinuity when the black hole shrinks to become a singular solution without horizon.

Thus for $k=0$, the CA and the new CV conjectures become the same. For $k=\pm 1$, on the other hand, we do not have a general equality.  Instead we propose the inequality in (\ref{ineq1}) and verify these using large classes of AdS black holes.

Our analysis also makes clear the condition for the Lloyd's bound $\dot I_{\rm WDW}\le 2M$.  The bound is always satisfied if we have positive volumes of the black hole singularities, but it can be violated when the volumes turn negative.  The question then naturally arises whether a black hole that satisfies the null energy condition can have negative singularity volume and hence violate the Lloyd's bound.  We shall address this issue in some great detail in section \ref{sec:negvol}.

\subsection{Differential ${\cal C}$-$S$ inequality}
\label{sec:difbound}

Based on our analysis of the Schwarzschild and RN AdS black holes, we also propose a differential inequality
\be
\fft{1}{T}\fft{\partial \dot I_{\rm WDW}}{\partial S}\Big|_{Q_i,P_{\rm th}} \ge C\,,\qquad \hbox{for} \qquad k=0,1,\label{diffineq}
\ee
where the entropy is understood as the standard one associated with the outer horizon and $C$ is a certain order-one numerical constant.  For the Schwarzschild-AdS black hole, we simply have
\be
\fft{1}{T}\fft{\partial \dot I_{\rm WDW}}{\partial S}\Big|^{\rm Sch-AdS}_{P_{\rm th}} =2\,,\qquad \hbox{for} \qquad k=0,1.\label{diffsch}
\ee
As we have demonstrated that this is the lower bound for the RN-AdS black holes.  However, we shall see in subsequent sections that the $C=2$ bound can be violated by other types of black holes. Nevertheless there exists a theory-dependent numerical constant $C$ of order one such that the inequality (\ref{diffineq}) holds. In this paper we shall examine large classes of $D\ge 4$ AdS black holes and we find in these examples that the lowest $C$ is $(D-3)/(D-2)$.

If instead we treat the quantity $2P_{\rm th}\, \Delta V_{\rm th}$ as the later-time growth rate of the holographic complexity, then we would like to propose the following inequality
\be
\fft{1}{T}\fft{\partial (2P_{\rm th}\,\Delta V_{\rm th})}{\partial S}\Big|_{Q,P_{\rm th}}
\qquad
\left\{
  \begin{array}{ll}
    >0\,, &\qquad \hbox{for}\qquad k=1\,; \\
    >C\,, &\qquad \hbox{for}\qquad k=0\,.
  \end{array}
\right.\label{diffineq2}
\ee
In the remaining sections, we shall go through large classes of AdS black holes in Einstein gravity with minimally coupled matter that satisfies the null energy condition and verify these equalities and inequalities.  In particular, we find that the Lloyd's bound and also the inequality (\ref{diffineq2}) for $k=1$ are not always satisfied. However, the other relations are surprisingly robust.

\section{Einstein-Maxwell-dilaton theories}
\label{sec:emd}

In this section, we consider Einstein-Maxwell-dilaton (EMD) types of theories that may involve multiple minimally coupled dilatonic scalars and multiple $U(1)$ Maxwell fields. These theories typically emerge from or are inspired by gauged supergravities and exact charged AdS black holes have been constructed and well studied, providing concrete examples to test our proposed inequalities. The scaling symmetries associated with constant shifts of dilatons typically breaks down by the scalar potential, but we shall still refer them as the EMD theories.

\subsection{$D=4$ gauged STU supergravity}
\label{sec:d4stu}

Four-dimensional ${\cal N}=2$ STU supergravity \cite{Duff:1995sm} is pure supergravity coupled to three vector multiplets.  The theory can be gauged with $U(1)^4$ gauge fields and it is a consistent truncation of ${\cal N}=8$ gauged supergravity.  For electrically-charged static black holes, the relevant Lagrangian involves four gauge fields $A_i$ and three dilatonic scalars.  The charged AdS black hole was constructed in \cite{Duff:1999gh} and its embedding in rotating M2-branes were given in \cite{Cvetic:1999xp}.  The solution is
\bea
ds_4^2 &=& -\prod_{i=1}^4 H_i^{-\fft12} \tilde f\, dt^2 +
\prod_{i=1}^4 H_i^{\fft12} \left( \fft{dr^2}{\tilde f} + r^2 d\Omega_{2,k}^2\right)\,,\nn\\
A^i &=& \fft{\sqrt{q_i (\mu + k q_i)}}{(r+q_i)} dt\,,\qquad
X_i = H_i^{-1} \prod_{j=1}^4 H_j^{\fft12}\,,
\eea
with
\be
\tilde f = 1 - \fft{\mu}{r} + g^2 r^2 \prod_{i=1}^4 H_i\,,\qquad H_i = 1 + \fft{q_i}{r}\,.
\ee
It contains five nonnegative integration constants $(\mu,q_1,q_2,q_3,q_4)$ parameterizing the mass and the four types of electric charges
\be
M=\ft12\mu + \ft14 k \sum_{j=1}^4 q_j\,,\qquad Q_i=\ft14\sqrt{q_i (\mu + k q_i)}\,,\qquad i=1,2,3,4.
\ee
Note we take the convention that the asymptotic region is $r\rightarrow +\infty$ and thus the reality condition requires $q_i\ge 0$. The curvature singularity is located at $r=-{\rm min}\{q_i\}$, which is zero if one of the charges vanishes. The horizon is located at $r=r_0$ with $\tilde f(r_0)=0$.  For $q_1q_2q_3q_4\ne 0$, we have two horizons $r_+\ge r_-> 0$. When $q_1q_2q_3q_4=0$, the black hole can have at most one horizon. We shall discuss the former case first and study the latter case later.  The thermodynamical variables related to the horizons are
\bea
T &=& \fft{\tilde f'}{4\pi} \prod_{j=1}^4 H_j^{-\fft12}\Big|_{r=r_0}\,,\qquad S=\pi \sqrt{\prod_{j=1}^4 (r_0+q_j)}\,,\nn\\
\Phi_i &=& \fft{\sqrt{q_i (\mu + k q_i)}}{r_0 + q_i}\,,\qquad
P_{\rm th} = \fft{3g^2}{8\pi}\,,\qquad V_{\rm th} =\ft13\pi r^3 \Big(\prod_{i=1}^4 H_i\Big)
\sum_{j=1}^4 H_i^{-1} \Big|_{r=r_0}\,.\label{Vthstu4}
\eea
We can verify the first law (\ref{genfl}).  We are now ready to discuss the holographic complexity bounds and we discuss different cases separately depending on the number of non-vanishing charges.

\bigskip
\noindent{\bf Case 1: $q_1 q_2 q_3 q_4\ne 0$}
\medskip

In this case, there are two horizons $r_+\ge r_->0$ and curvature singularity is located at $r=0$. (If the cosmological constant vanishes, $r=0$ is in fact the inner horizon.) Thus we have
\be
\dot I_{\rm WDW} = \sum_{j=1}^4\Phi_j Q_j\Big|^{r_-}_{r_+}\,.
\ee
It is straightforward to establish the identities
\bea
&&\dot I_{\rm WDW} = 2P_{\rm th}\, \Delta V_{\rm th} + k (r_+-r_-)\,,\nn\\
&&\dot I_{\rm WDW} - 2M\nn\\
&&=-\ft14 \mu \sum_i \Big(\fft{r_-}{r_-+q_i} +
\fft{q_i}{r_+ + q_i}\Big) - \ft14k \sum_i\Big(\fft{q_i (2r_-+q_i)}{r_-+q_i} +
\fft{q_i^2}{r_+ + q_i}\Big)\,.
\eea
Note that for $k=0,1$, we must have $\mu>0$. It follows that the inequalities (\ref{ineq1}) must be satisfied.

The differential bound (\ref{diffineq}) can also be established and we find $C=4/3$. However the detail can be rather messy and we shall give the description of the procedure.  We can first express parameters $(\mu, q_4)$ in terms of inner and outer horizons $(r_-,r_+)$. Keeping the charges $(Q_1,Q_2,Q_3,Q_4)$ fixed implies that the parameters $(q_1,q_2,q_3,r_-)$ can all be expressed in terms of $r_+$ and we can obtain $(q_1', q_2', q_3', r_-')$, where a prime denotes a derivative with respect to $r_+$.  It is then straightforward to calculate the quantity
\be
Z= \fft{1}{T} \fft{\partial \dot I_{\rm WDW}}{\partial S}\Big|_{Q_i, P_{\rm th}} - \fft43\,.
\ee
We find that $Z$ is a rational polynomial of $(q_1,q_2,q_3, g^2, k)$ involving thousands of terms.  We need to impose the reality condition for $Q_i$, namely
\be
q_1 q_2 q_3-\left(q_1+q_2+q_3\right) r_- r_+-r_- r_+ \left(r_-+r_+\right)\equiv\alpha>0\,.
\ee
We can substitute $q_1 q_2 q_3$ into $Z$ in appropriate way, then we find that $Z$ becomes a rational polynomials of $(q_i, g^2, k, \alpha)$ with all positive coefficients and hence it is nonnegative for $k=0,1$.

\bigskip
\noindent{\bf Case 2: $q_3=q_4=0$}
\medskip

We now consider the simpler case with two vanishing charges. Without loss of generality we set $q_3=q_4=0$.  The parameters $q_1$ and $q_2$ are free.  In this case, there is only one horizon $r_+>0$, related to $\mu$ by
\be
\mu=r_+\big(k + g^2(r_++q_1)(r_++q_2)\big)\,.
\ee
In other words, for given $q_1$ and $q_2$, the horizon radius can shrink to zero, giving rise to spacetime with naked singularity. We find that the growth rate of the action in the WDW patch is
\be
\dot I_{\rm WDW} = 2M - (\Phi_1 Q_1 + \Phi_2 Q_2) - \ft14 k (q_1 + q_2)\,.
\ee
Note that in this case we have $V_{\rm th}^0=0$.  It is then easy to establish the identity
\bea
\dot I_{\rm WDW} &=& 2 P_{\rm th} V_{\rm th}^+ + k r_+\nn\\
&=& 2 M -\ft12 k(q_1 + q_2) - \ft14 g^2 r_+ (2q_1 q_2 + q_1 r_+ + q_2 r_+)\,.
\eea
Thus the algebraic inequalities presented in (\ref{ineq1}) are all satisfies, with the understanding that $\Delta V_{\rm th}=V_{\rm th}^+$ since there is only one horizon and $V_{\rm th}^0=0$.

The differential relation (\ref{diffineq}) in this case can also be easily established.  For $k=0$, we find
\bea
&& \fft{1}{T} \fft{\partial I_{\rm WDW}}{\partial S}\Big|_{Q_i,P_{\rm th}} -1\nn\\
&=&\frac{r_+ \left(11 \left(q_1+q_2\right) r_+^2+2 \left(q_1^2+11 q_2 q_1+q_2^2\right) r_++5 q_1 q_2 \left(q_1+q_2\right)+6 r_+^3\right)}{2 \left(q_1+r_+\right) \left(q_2+r_+\right) \left(2 \left(q_1+q_2\right) r_++q_1 q_2+3 r_+^2\right)}>0\,.
\eea
For $k=1$, the expression becomes more complicated and it is not worth giving explicitly here.  We find that the expression is still a rational polynomials of $(r_+,q_1,q_2,g^2)$ with the coefficients of polynomial terms all positive, and hence the differential inequality (\ref{diffineq}) is also satisfied, with $C=1$. If we set further $q_2=0$, we find now $C=\ft32$.

\bigskip
\noindent{\bf Case 3: $q_4=0$, $q_1 q_2 q_3\ne 0$}
\medskip

We now consider the case where there is only one vanishing charge and we set $q_4=0$, with $q_1q_2q_3\ne 0$.  There is also only one real horizon $r_+>0$, satisfying
\be
\mu=g^2 q_1 q_2 q_3 + \Big(k + g^2(q_1 q_2 + q_2 q_3 + q_1 q_3)\Big)r_+ +
g^2 (q_1 + q_2 + q_3) r_+^2 + g^2 r_+^3\,.
\ee
Intriguingly, when $\mu=g^2 q_1 q_2 q_3$, corresponding to $r_+=0$, the $r=0$ region gives rise to a singular spacetime that is conformal to AdS$_2\times \Omega_{2,k}$. The action growth rate is
\be
\dot I_{\rm WDW} = 2M - (\Phi_1 Q_1 +\Phi_2 Q_2 + \Phi_3 Q_3) - \ft14 k (q_1 + q_2 + q_3) - 2 P_{\rm th} V_{\rm th}^0\,,
\ee
where $V_{\rm th}^0$ is the volume of the singularity $r=0$, given by
\be
V_{\rm th}^0 = \ft13 \pi q_1 q_2 q_3 >0\,.
\ee
It is then straightforward to verify that
\be
\dot I_{\rm WDW} = 2P_{\rm th} \Delta V_{\rm th} + k r_+\,,\qquad  \Delta V_{\rm th}=V_{\rm th}(r_+) - V_{\rm th}^0\,.
\ee
Thus we see that the inequalities (\ref{ineq1}) are all satisfied.

The differential bound (\ref{diffineq}) can also be established. For $k=0$, we have
\bea
&&\fft{1}{T} \fft{\partial \dot I_{\rm WDW}}{\partial S}\Big|_{Q_i,P_{\rm th}}=1 + \fft{X}{Y}\,,\nn\\
X &=&6 r_+^5+11 \left(q_1+q_2+q_3\right) r_+^4+2 \left(q_1^2+11 \left(q_2+q_3\right) q_1+q_2^2+q_3^2+11 q_2 q_3\right) r_+^3\nn\\
&&+\left(5 \left(q_2+q_3\right) q_1^2+\left(5 q_2^2+57 q_3 q_2+5 q_3^2\right) q_1+5 q_2 q_3 \left(q_2+q_3\right)\right) r_+^2\nn\\
&&+20 q_1 q_2 q_3 \left(q_1+q_2+q_3\right) r_++6 q_1 q_2 q_3 \left(q_2 q_3+q_1 \left(q_2+q_3\right)\right)\,,\\
Y&=&2 \left(q_1+r_+\right) \left(q_2+r_+\right) \left(q_3+r_+\right)
\left(3 r_+^2+ 2 \left(q_1+q_2+q_3\right) r_++q_2 q_3+q_1 \left(q_2+q_3\right)\right)\,.\nn
\eea
The expression for $k=1$ is much more complicated, but the same inequality (\ref{diffineq}) holds with $C=1$.

We would like to clarify the concept of the volume $V_{\rm th}^0$ here with this concrete example. It appears in the expression of $\dot I_{\rm WDW}$ such that the quantity vanishes when $r_+\rightarrow 0$, for which the solution ceases to be a black hole.  To understand this quantity, we compare the thermodynamical volume with the entropy of the black hole. Although the black hole entropy is one quarter of the area of the horizon, the concept of area of the foliating sphere can be locally defined at any $r$.  Analogously, although we derive the thermodynamical volume using the horizon properties, the concept of volume may be locally defined at any $r$, and it becomes the thermodynamical volume when $r=r_+$. In fact, for a class of static black holes, a local volume formula was given in \cite{Feng:2017wvc}. In this explicit example, we can generalize the $V_{\rm th}(r_0)$ in (\ref{Vthstu4}) defined on the horizon $r_0$ to any $r$, namely \cite{Feng:2017wvc}
\be
V_{\rm th}(r) =\ft13\pi r^3 \Big(\prod_{j=1}^4 H_i(r)\Big)
\sum_{j=1}^4 \fft{1}{H_i(r)}\,.
\ee
It can be argued that the name ``volume'' is not suitable for $V_{\rm th}(r)$ in the WDW patch since $r$ is timelike; however, we shall continue to use the terminology for lacking a better description. It is important to note that this quantity is independent of $g^2$ hence the pressure $P_{\rm th}$. Therefore the concept of volume can still be valid for asymptotically flat black holes with $g=0$ \cite{Feng:2017wvc}. For general $q_i$ parameters, we can choose, without loss of generality, that $q_4\le q_3\le q_2\le q_1$, then the curvature singularity is located at $r=-q_4$.  The volume of the singularity is thus
\be
V_{\rm th}^0= V_{\rm th}(-q_4) =
\ft43 \pi (q_1-q_4)(q_2-q_4)(q_3-q_4)\ge 0\,.
\ee
When the charge parameters are all equal, the solution becomes the RN black hole and we recovers $V_{\rm th}^0=0$.  In general when $q_1q_2q_3q_4\ne 0$, the black hole has two horizons and the curvature singularity is excluded from the WDW patch and hence $V_{\rm th}^0$ has no direct effect on our results. On the other hand, when $q_4=0$, there is only a single horizon such that the curvature singularity becomes part of the WDW patch, then the non-vanishing $V_{\rm th}^0$ will have nontrivial contributions. In section \ref{sec:negvol}, we shall present examples of single-horizon black holes where $V_{\rm th}^0$ is negative or even divergent.

We now examine the inequality (\ref{diffineq2}). For $k=0$, we simply have $\dot I_{\rm WDW}=2P_{\rm th}\, \Delta V_{\rm th}$. For $k=1$, we find that the inequality (\ref{diffineq2}) holds for general parameters.  The general expression is very complicated, and as a concrete representative example, we consider the case 2 discussed above and find
\be
\fft{1}{T}\fft{\partial (2 P_{\rm th} \Delta V_{\rm th})}{\partial S}\Big|_{Q_1, P_{\rm th}} = \fft{g^2 X}{Y}>0\,,
\ee
where the expressions for $X$ and $Y$ are quite complicated even in this simple solution, given by
\bea
X&=&12 r_+^4+ 27 \left(q_1+q_2\right) r_+^3+2 \left(6 q_1^2+29 q_2 q_1+6 q_2^2\right) r_+^2\nn\\
&&+24 q_1 q_2 \left(q_1+q_2\right) r_++8 q_1^2 q_2^2 + g^2 r_+\Bigg(24 r_+^5+ 60 \left(q_1+q_2\right) r_+^4\nn\\
&&+\left(45 q_1^2+138 q_2 q_1+45 q_2^2\right) r_+^3+4 \left(3 q_1^3+22 q_2 q_1^2+22 q_2^2 q_1+3 q_2^3\right) r_+^2\nn\\
&&+4 q_1 q_2 \left(5 q_1^2+9 q_2 q_1+5 q_2^2\right) r_++4 q_1^2 q_2^2 \left(q_1+q_2\right)
\Bigg)\nn\\
&&+ g^2 (r_+ + q_1)(r_+ + q_2) \Big(12 r_+^4+21 \left(q_1+q_2\right) r_+^3+\left(6 q_1^2+38 q_2 q_1+6 q_2^2\right) r_+^2\cr
&&+11 q_1 q_2 \left(q_1+q_2\right) r_++2 q_1^2 q_2^2\Big)\,,\nn\\
Y&=&\Big(1+g^2 \left(3 r_+^2+2 \left(q_1+q_2\right) r_++q_1 q_2\right)\Big)\Bigg(2 r_+^2+ 3 \left(q_1+q_2\right) r_++4 q_1 q_2\nn\\
&&+ g^2\Big(r_+ \left(q_1+r_+\right) \left(q_2+r_+\right) \left(4 r_++3 \left(q_1+q_2\right)\right)\Big)\nn\\
&&+ 2g^4 r_+^2 \left(q_1+r_+\right){}^2 \left(q_2+r_+\right)^2\Bigg)\,.
\eea
Since $X$ and $Y$ are manifestly positive, the differential inequality (\ref{diffineq2}) holds for electrically charged black holes in $D=4$ gauged STU supergravity.

     For the differential relation (\ref{diffineq}), we have focused on the discussion for $k=0,1$.
Before finishing this subsection, we would like to present an example for the $k=-1$ case.  The simplest solution is to set $q_2=q_3=q_4=0$.  For $q_1>0$, the reality condition requires that
\be
\mu-q_1=\alpha\ge 0\,,
\ee
which ensures that the black hole has one single horizon $r_+$, with well defined global structure and thermodynamical variables.  We find
\bea
&&\fft{1}{T}\fft{\partial \dot I_{\rm WDW}}{\partial S}\Big|_{Q_1, P_{\rm th}}\\
&&=\frac{6 q_1^2 \left(\alpha ^2-q_1^2\right)+3 r_+^2 \left(4 \alpha ^2+5 \alpha  q_1-6 q_1^2\right)+q_1 r_+ \left(21 \alpha ^2+7 \alpha  q_1-18 q_1^2\right)+r_+^3 \left(8 \alpha -6 q_1\right)}{\left(2 \alpha -q_1\right) \left(q_1+r_+\right) \left(2 q_1 \left(\alpha +q_1\right)+r_+ \left(3 \alpha +4 q_1\right)+2 r_+^2\right)}.\nn
\eea
This quantity can be negative.  For example, let $\alpha=3q_1/5$, we have
\be
\fft{1}{T}\fft{\partial \dot I_{\rm WDW}}{\partial S}\Big|_{Q_1, P_{\rm th}}=
-\frac{3 \left(39 q_1 r_+^2+52 q_1^2 r_++32 q_1^3+10 r_+^3\right)}{\left(q_1+r_+\right) \left(29 q_1 r_++16 q_1^2+10 r_+^2\right)} <0\,.
\ee
We shall thus not consider the $k=-1$ case for the differential inequalities in the remainder of the paper.

\subsection{$D=5$ $U(1)^3$ gauged supergravity}
\label{sec:d5stu}

The $U(1)^3$ gauged supergravity in five dimensions has two vector multiplets and it can be consistently constructed from maximum gauged supergravity.  The bosonic sector consists of the metric, three $U(1)$ gauged fields $A_i$ and two dilatonic scalars.  The asymptotically AdS solution is given by \cite{Behrndt:1998jd,Cvetic:1999xp}
\bea
ds_5^2 &=& (H_1H_2H_3)^{-\fft23} \tilde f dt^2 +
(H_1 H_2 H_3)^{\fft13} \Big(\fft{dr^2}{\tilde f} + r^2 d\Omega_{3,k}^2\Big)\,,\nn\\
A^i &=& \fft{\sqrt{q_i (\mu + k q_i)}}{(r^2+q_i)} dt\,,\qquad
X_i = H_i^{-1} \prod_{j=1}^4 H_j^{\fft12}\,,
\eea
with
\be
\tilde f = 1 - \fft{\mu}{r} + g^2 r^2 \prod_{i=1}^3 H_i\,,\qquad H_i = 1 + \fft{q_i}{r^2}\,.
\ee
The solution contains four integration constants $(\mu, q_1,q_2,q_3)$, parameterizing the mass and electric charges
\be
M=\ft{3}8\pi\Big(\mu + \ft23k (q_1 + q_2 + q_3)\Big)\,,\qquad
Q_i = \ft14\pi \sqrt{q_i (\mu + k q_i)}\,.
\ee
For appropriate parameters, the solution describes a black hole with the horizon $r_0$ with $f(r_0)=0$, namely
\be
\mu=k r_0^2 + g^2 r_0^{-2} (r_0^2 + q_1) (r_0^2+q_2) (r_0^2 + q_3)\,.
\ee
The rest of thermodynamical variables are
\bea
T &=& \fft{f'}{4\pi} \prod_{j=1}^3 H_j^{-\fft12} \Big|_{r=r_0}\,,\qquad
S=\ft12 \pi^2 \sqrt{\prod_{j=1}^3 (r_0^2 + q_j)}\,,\nn\\
\Phi_i &=& \fft{\sqrt{q_i(\mu + k q_i)}}{r_0^2 + q_i}\,,\qquad P_{\rm th}=\fft{3g^2}{4\pi}\,,\qquad
V_{\rm th} = \ft16\pi^2 r^4 \Big(\prod_{i=1}^3 H_i\Big) \sum_{j=1}^3 H_j^{-1}\Big|_{r=r_0}\,.
\eea
As was discussed in the end of the previous subsection, we can define a local $V_{\rm th}$ that is a function of general $r$ rather than only $r=r_0$. Furthermore, the volume is independent of the effective cosmological constant or the thermodynamical pressure. This generalized volume formula allows us to discuss the volume at the curvature singularity. We are now in the position to study the holographic complexity and we analyse it case by case depending on the charge configuration.

\bigskip
\noindent{\bf Case 1: $q_1q_2q_3\ne 0$}
\medskip

There are two horizons $r_+\ge r_->0$ when all charges are turned on. The action growth rate in the WDW patch is
\be
\dot I_{\rm WDW}=\sum_{j=1}^3 \Phi_j Q_j \Big|_{r_+}^{r_-}\,.
\ee
As in the four-dimensional STU model, it is also straightforward to establish the identities
\bea
&&\dot I_{\rm WDW} = 2 P_{\rm th} \Delta V_{\rm th} + \ft34\pi k (r_+^2 - r_-^2)\,,\nn\\
&&\dot I_{\rm WDW} - 2M\nn\\
&&=-\ft14\pi \mu \sum_{i=1}^3 \Big(\fft{r_-^2}{r_-^2+q_i} +
\fft{q_i}{r_+^2 + q_i}\Big) - \ft14\pi k \sum_{i=1}^3 \Big(\fft{q_i (2r_-^2+q_i)}{r_-^2+q_i} +
\fft{q_i^2}{r_+^2 + q_i}\Big)\,.
\eea
Note that for $k=0,1$, we must have $\mu>0$. It follows that the relations in (\ref{ineq1}) must be all satisfied.

The analysis for the differential bound is much simpler than the previous four-charge STU model in four dimensions and can be analysed analytically for general parameters.  However, for simplicity of the presentation, we consider a special case $q_2=q_1$ and hence $Q_2=Q_1$. We can solve for $(\mu,q_3)$ in terms of $(r_+,r_-)$. We find that the reality condition requires that
\be
 q_1-r_+ r_- =\alpha>0\,.
\ee
Keeping the charges $Q_i$ and the pressure $P_{\rm th}$ fixed, we find
\bea
&&\fft{1}{T} \fft{\partial \dot I_{\rm WDW}}{\partial S}\Big|_{Q_i, P_{\rm th}} = \fft43 + \fft{X}{Y}\,,\nn\\
X&=& 8 \alpha ^8 \left(2 r_-^2+r_+^2\right)+2 \alpha ^7 \left(15 r_-^4+64 r_+ r_-^3+40 r_+^2 r_-^2+32 r_+^3 r_-+5 r_+^4\right)\nn\\
&&+\alpha ^6 \left(21 r_-^6+210 r_+ r_-^5+559 r_+^2 r_-^4+560 r_+^3 r_-^3+309 r_+^4 r_-^2+70 r_+^5 r_-+5 r_+^6\right)\nn\\
&&+\alpha ^5 \left(r_-+r_+\right){}^2 (5 r_-^6+116 r_+ r_-^5+453 r_+^2 r_-^4+540 r_+^3 r_-^3\nn\\
&&+195 r_+^4 r_-^2+28 r_+^5 r_-+r_+^6)+\alpha ^4 r_- r_+ \left(r_-+r_+\right){}^2\nn\\
&&\times\left(25 r_-^6+272 r_+ r_-^5+781 r_+^2 r_-^4+926 r_+^3 r_-^3+407 r_+^4 r_-^2+74 r_+^5 r_-+5 r_+^6\right)\nn\\
&&+2 \alpha ^3 r_-^2 r_+^2 \left(r_-+r_+\right){}^4 \left(24 r_-^4+128 r_+ r_-^3+149 r_+^2 r_-^2+48 r_+^3 r_-+5 r_+^4\right)\nn\\
&&+2 \alpha ^2 r_-^3 r_+^3 \left(r_-+r_+\right){}^4 \left(22 r_-^4+84 r_+ r_-^3+87 r_+^2 r_-^2+39 r_+^3 r_-+5 r_+^4\right)\nn\\
&&+2 \alpha  r_-^4 r_+^4 \left(r_-+r_+\right){}^6 \left(9 r_-^2+10 r_+ r_-+2 r_+^2\right)+2 r_-^6 r_+^5 \left(r_-+r_+\right){}^6 \left(r_-+2 r_+\right), \nn\\
Y&=&\ft34 (r_+^2-r_-^2) (r_-^2 + q_1)^2 (r_+^2 + q_1)^2 \left(2 \alpha ^2+\alpha  r_- \left(r_-+4 r_+\right)+r_-^2 r_+ \left(r_-+r_+\right)\right)\nn\\
&&\times\left(2 \alpha ^2+\alpha  r_+ \left(4 r_-+r_+\right)+r_- r_+^2 \left(r_-+r_+\right)\right).
\eea
Thus we see that the bound (\ref{diffineq}) is satisfied with $C=4/3$.  The bound is saturated in the limit of $r_+\rightarrow \infty$, while keeping $(q_1,r_-)$ fixed.  This is however not the Schwarzschild limit, which is achieved by setting $r_-=0$ first and then set $\alpha=0$. For the Schwarzschild-AdS limit, we have (\ref{diffsch}). For the charged AdS black holes in the $U(1)^3$ theory, the bound is lower at $C=4/3$.

\bigskip
\noindent{\bf Case 2: $q_2=q_3=0$}
\medskip

In this case, there is only one horizon $r_+$, given by
\be
\mu=r_+^2 \big(k + g^2 (r_+^2 + q_1)\big)\,.
\ee
The growth rate of the action in the WDW patch is
\bea
\dot I_{\rm WDW} &=&  2 M - \Phi_1 Q_1 - \ft14 \pi k q_i\nn\\
&=& 2 P_{\rm th} V_{\rm th} + \ft34\pi k r_+^2\nn\\
&=& 2M - \ft14 \pi q_1 (2k + g^2 r_+^2)\,.
\eea
Note that we have $V_{\rm th}^0=0$ in this case. It follows that the inequality relations (\ref{ineq1}) are satisfied.  Keeping $Q_1$ and $P_{\rm th}$ fixed, we find
\bea
&&\fft{1}{T} \fft{\partial \dot I_{\rm WDW}}{\partial S}\Big|_{Q_1, P_{\rm th}} - \fft43\nn\\
&=& \fft{2 \left(6 g^4 r_+^4 \left(2 q_1+r_+^2\right)+g^2 k \left(26 q_1 r_+^2+4 q_1^2+9 r_+^4\right)+k^2 \left(10 q_1+3 r_+^2\right)\right)}{3(k +  g^2 (2r_+^2 + q_1))\left(3 g^2 r_+^2 \left(q_1+r_+^2\right)+k \left(4 q_1+3 r_+^2\right)
\right)}\,.
\eea
Thus the bound (\ref{diffineq}) is satisfied with again $C=\fft43$.

\bigskip
\noindent{\bf Case 3: $q_3=0$ and $q_1 q_2\ne 0$}
\medskip

In this case, there is also only just one horizon $r_+$, related to $\mu$ by
\be
\mu=g^2 \left(q_1+r_+^2\right) \left(q_2+r_+^2\right)+k r_+^2\,.
\ee
The action growth rate is
\bea
\dot I_{\rm WDW} &=& 2M - (\Phi_1 Q_1 + \Phi_2 Q_2)-2P_{\rm th} V_{\rm th}^0 - \ft14\pi k (q_1 + q_2)\nn\\
&=& 2M -\frac{1}{4} \pi  \left(2 k \left(q_1+q_2\right) + g^2 \left(q_1 r_+^2+q_2 r_+^2+3 q_1 q_2\right)
\right)\nn\\
&=&2 P_{\rm th}\,\Delta V_{\rm th} + \ft34\pi k r_+^2\,.
\eea
Note that in this case, we have $2 P_{\rm th} V_{\rm th}^0= \ft14 \pi g^2 q_1 q_2$. The relation (\ref{ineq1}) are again satisfied.

Keeping $(Q_1,Q_2,P_{\rm th})$ fixed, we find that
\be
\fft{1}{T} \fft{\partial \dot I_{\rm WDW}}{\partial S}\Big|_{Q_i,P_{\rm th}} = \fft43 + X\,,
\ee
where $X$ is an explicit positive quantity for $k=0,1$.  For $k=0$, the expression of $X$ is fairly simple, and we present it here:
\be
X=\frac{4 \left(2 \left(q_1+q_2\right) r_+^4+7 q_1 q_2 r_+^2+2 q_1 q_2 \left(q_1+q_2\right)+r_+^6\right)}{3 \left(q_1+r_+^2\right) \left(q_2+r_+^2\right) \left(q_1+q_2+2 r_+^2\right)}\,.
\ee
It follows that the bound (\ref{diffineq}) is satisfied with $C=4/3$.

Finally we would like to mention that we have also verified the inequality (\ref{diffineq2}) holds for all black hole parameters.  As a concrete example, we present the result for the Case 2 with $k=1$. In this case, there is only one horizon and $V_{\rm th}^0=0$.  We find
\be
\fft{1}{T} \fft{\partial (2P_{\rm th}\,\Delta V_{\rm th})}{\partial S}\Big|_{Q_1, P_{\rm th}} =
\fft{4g^2\Big(g^2 r_+^2 \left(3 r_+^4+5 q_1 r_+^2+q_1^2\right)+3 r_+^4+6 q_1 r_+^2+2 q_1^2\Big)}{\Big(1+g^2 \left(q_1+2 r_+^2\right)\Big)\Big(3 g^2 r_+^2 \left(r_+^2+q_1\right)+3 r_+^2+4 q_1\Big)}>0\,.
\ee

\subsection{$U(1)^2$ theory-I in general $D$ dimensions}
\label{sec:chow}

The $D=4,5$ gauged supergravities with two non-vanishing charges were generalized to general higher dimensions \cite{Chow:2011fh}. The relevant Lagrangian consists of two dilatonic scalars and two $U(1)$ gauged fields. Turing off the gauging, the two charges are associated with the Kaluza-Klein and winding modes of the bosonic strings. The theory can be also embedded in gauged supergravities in $D=6,7$. The solution in general dimensions is \cite{Chow:2011fh}
\bea
ds_D^2 &=& (H_1 H_2)^{\fft1{D-2}} \Big(-\fft{f}{H_1 H_2} dt^2 + \fft{dr^2}{f} + r^2 d\Omega_{D-2,k}\Big)\,,
\nn\\
A_i &=& \fft{\sqrt{q_i(\mu + k q_i)}}{r^{D-3} + q_i} dt\,,\qquad X_i= H_i^{-1} (H_1 H_2)^{\fft{D-3}{2(D-2)}}\,,\nn\\
f &=& k - \fft{\mu}{r^{D-3}} + g^2 r^2 H_1 H_2\,,\qquad H_i = 1 + \fft{q_i}{r^{D-3}}\,.
\eea
The solution contains three integration constants $(\mu, q_1, q_2)$ parameterizing the mass and two electric charges
\be
M=\fft{\Omega_{D-2}}{16\pi} \Big((D-2)\mu + (D-3) k (q_1 + q_2)\Big)\,,\qquad
Q_i = \fft{(D-3)\Omega_{D-3,k}}{16\pi} \sqrt{q_i (\mu + k q_i)}\,.
\ee
The solution describes a black hole when there exists a positive $r_0$ such that $f(r_0)=0$.
The remaining thermodynamical variables are
\bea
T &=& \fft{f'}{4\pi \sqrt{H_1 H_2}}\Big|_{r=r_0}\,,\qquad S=\ft14 \Omega_{D-2} r^{D-2} \sqrt{H_1 H_2}
\Big|_{r=r_0}\,,\nn\\
\Phi_i &=& \fft{\sqrt{q_i (\mu + k q_i)}}{r_0^{D-3} + q_i}\,,\qquad
P_{\rm th} = \fft{(D-1)(D-2)g^2}{16\pi}\,,\nn\\
V_{\rm th} &=& \fft{\Omega_{D-2} r^{D-1}}{2(D-1)(D-2)}
\Big(2 H_1 H_2 + (D-3) (H_1 + H_2)\Big)\Big|_{r=r_0}\,.
\eea
The $D=4,5$ cases were included in the previous subsections and these black holes have only one horizon.  The $D=6,7$ solutions were also given in
\cite{Cvetic:1999un} and \cite{Cvetic:1999xp} respectively. We shall focus on $D\ge 6$.

\bigskip
\noindent{\bf Case 1: $q_1 q_2\ne 0$}
\medskip

When both charges are non-vanishing, there are two horizons $r_+\ge r_->0$ for dimensions $D\ge 6$.  The action growth rate is given by
\bea
\dot I_{\rm WDW} &=& (\Phi_1 Q_1 + \Phi_2 Q_2)\Big |_{r_+}^{r_-}\nn\\
&=& 2 P_{\rm th}\, \Delta V_{\rm th} + \fft{(D-2)\Omega_{D-2}}{8\pi}\,k\Big(r_+^{D-3} - r_-^{D-3}\Big)\nn\\
&=& 2M + \fft{\Omega_{D-2}}{16\pi}(D-3)\Bigg[\mu \Big(\fft{2}{D-3} +\sum_{i=1}^2 \big(\fft{r_-^{D-3}}{r_-^{D-3}+q_i} +
\fft{q_i}{r_+^{D-3} + q_i}\big)\Big)\nn\\
&&\qquad +k \sum_{i=1}^2\Big(\frac{q_i^2}{r_+^{D-3}+q_i}+\frac{q_i (2 r_-^{D-3}+q_i)}{r_-^{D-3}+q_i}
\Big)
\Bigg]\,.
\eea
It follows that the inequalities in (\ref{ineq1}) are satisfied.  For the differential bound (\ref{diffineq}), we find that for $D\ge 6$, we have
\be
C = \fft{D-3}{D-2}\,, \qquad D\ge 6\,.
\ee
We now present the case explicitly for $D=7$.  We can solve for $(\mu,q_2)$ in terms of $(q_1,r_-,r_+)$.  The reality condition of the solution then requires that
\be
\alpha=q_1- r_-^2 r_+^2> 0\,.
\ee
Keeping $(Q_1,Q_2,P_{\rm th})$ fixed, we find
\be
\fft{1}{T} \fft{\partial \dot I_{\rm WDW}}{\partial S}\Big|_{Q_i, P_{\rm th}}=\ft45 + Z\,,\qquad \hbox{for}\qquad
D=7\,,
\ee
where $Z$ is manifestly positive.  For $k=1$, the expression is too long to present there; we can however give the $k=0$ expression explicitly, namely $Z=X/Y$, with
\bea
X &=& 2 \alpha ^6 \left(r_-^2+4 r_+^2\right) r_-^2+2 \alpha ^5 \left(r_-^2+r_+^2\right)^2 \left(4 r_-^2+r_+^2\right) \left(2 r_-^2+3 r_+^2\right)\nn\\
&&+\alpha ^4 \left(r_-^2+r_+^2\right)^2 \left(23 r_-^8+108 r_+^2 r_-^6+144 r_+^4 r_-^4+66 r_+^6 r_-^2+9 r_+^8\right)\nn\\
&&+\alpha ^3 \left(r_-^2+r_+^2\right)^4 \left(9 r_-^8+80 r_+^2 r_-^6+128 r_+^4 r_-^4+40 r_+^6 r_-^2+3 r_+^8\right)\nn\\
&&+\alpha ^2 r_+^2 r_-^2\left(r_-^2+r_+^2\right)^4 \left(23 r_-^8+108 r_+^2 r_-^6+144 r_+^4 r_-^4+66 r_+^6 r_-^2+9 r_+^8\right)\nn\\
&&+2 \alpha  r_+^4 r_-^4\left(r_-^2+r_+^2\right)^6 \left(4 r_-^2+r_+^2\right) \left(2 r_-^2+3 r_+^2\right) \nn\\
&&+2 r_+^6 r_-^8\left(r_-^2+r_+^2\right)^6 \left(r_-^2+4 r_+^2\right),\cr
Y &=& \ft54 (r_+^2 - r_-^2) (q_1 + r_-^4) (q_1 + r_+^4)\nn\\
&&\times \left(\alpha ^2+\alpha  \left(r_-^2+r_+^2\right) \left(3 r_-^2+r_+^2\right)+r_-^2 r_+^2 \left(r_-^2+r_+^2\right)^2\right)\nn\\
&&\times \left(\alpha ^2+\alpha  \left(r_-^2+r_+^2\right) \left(r_-^2+3 r_+^2\right)+r_-^2 r_+^2 \left(r_-^2+r_+^2\right)^2\right).
\eea

We now examine the inequality (\ref{diffineq2}) and we find counter examples in $D\ge 6$ dimensions for $k=1$.  We present one explicit example in $D=7$, with following parameters
\be
g=\ft{1}{10}\,,\quad \mu=\ft{856639}{31750}\,,\quad q_1=100\,,\qquad q_2=\ft{8008}{127}\,,
\ee
such that $r_-=1$ and $r_+=2$.  We find that
\be
\fft{1}{T} \fft{\partial (2P_{\rm th}\,\Delta V_{\rm th})}{\partial S}\Big|_{Q_i, P_{\rm th}}=
-\ft{9254336430089727344731093}{14136367665344334011865000}\sim -0.655\,.
\ee
It should be pointed out that this violation will not occur for this class of solutions in $D=4,5$, where there can only be one horizon instead and the analysis is completely different, as was demonstrated in the STU models.  The violation is somewhat surprising since these black holes can be embedded in $D=7$ gauged supergravity that has the M-theory origin.

\bigskip
\noindent{\bf Case 2: $q_2= 0$}
\medskip

In this case, there is only one horizon $r_+$, satisfying
\be
\mu = k r_+^{D-3}+ g^2 r_+^2 (r_+^{D-3} + q_1)\,.
\ee
We find that the growth rate of the action in the WDW patch is
\bea
\dot I_{\rm WDW} &=& 2M - \Phi_1 Q_1 -\fft{(D-3)\Omega_{D-2}}{16\pi} k q_1\nn\\
&=& 2M - \fft{(D-3)\Omega_{D-2}}{16\pi} q_1 (2k + g^2 r_0^2)\nn\\
&=& 2P_{\rm th} V_{\rm th} + \fft{(D-2)\Omega_{D-2}}{8\pi} k r_+^{D-3}\,.
\eea
It is thus clear the relations in (\ref{ineq1}) are all satisfied.  Note that $V_{\rm th}^0=0$ in this case. For the differential bound (\ref{diffineq}), the example is sufficiently simple, we can present the full results.  Keeping $(Q_1,P)$ fixed, we find
\be
\fft{1}{T} \fft{\partial \dot I_{\rm WDW}}{\partial S}\Big|_{Q_1,P_{\rm th}} = \fft{D-1}{D-2} + \fft{X}{Y}\,,
\ee
where
\bea
X&=& g^2 k r_+^2 \left(\left(5 D^2-18 D+17\right) q_1 r_+^{D+3}+2 (D-1) q_1^2 r_+^6+2 (D-2)^2 r_+^{2 D}\right)\nn\\
&&+(D-2) (D-1) g^4 r_+^{D+4} \left(r_+^D+2 q_1 r_+^3\right)\nn\\
&&+(D-3) k^2 r_+^D \left((3 D-5) q_1 r_+^3+(D-2) r_+^D\right)\,,\nn\\
Y&=& \fft{D-2}{D-3} \Big(g^2 r_+^2 \left((D-1) r_+^D+2 q_1 r_+^3\right)+(D-3) k r_+^D\Big)\nn\\
&&\times \Big((D-2) g^2 r_+^2 \left(r_+^D+q_1 r_+^3\right)+k \left((D-1) q_1 r_+^3+(D-2) r_+^D\right)\Big)\,.
\eea
It is clear that $X/Y$ is manifestly positive for $k=0,1$ and the bound is $C=(D-1)/(D-2)$.

The differential inequality (\ref{diffineq2}) can also be easily established.  We find
\be
\fft{1}{T} \fft{\partial (2P_{\rm th}\, \Delta V_{\rm th})}{\partial S}\Big|_{Q_1,P_{\rm th}}=(D-1) g^2 r_+^2\, \fft{X}{Y}>0\,,
\ee
where $X$ and $Y$ are given by
\bea
X&=&g^2 r_+^2 \left((3 D-5) q_1 r_+^{D+3}+2 (D-2) r_+^{2 D}+2 q_1^2 r_+^6\right)\nn\\
&&+3 (D-1) q_1 r_+^{D+3}+2 (D-2) r_+^{2 D}+4 q_1^2 r_+^6\,,\nn\\
Y&=&\Big(g^2 r_+^2 \left((D-1) r_+^D+2 q_1 r_+^3\right)+(D-3) r_+^D\Big)\nn\\
&&\times \Big((D-2) g^2 r_+^2 \left(r_+^D+q_1 r_+^3\right)+(D-1) q_1 r_+^3+(D-2) r_+^D\Big)\,.
\eea
This result is not surprising since there is only one horizon for all $D\ge 4$ and we have already established in the STU models that this bound is valid for $D=4,5$ and thus its generalization to higher dimensions is expected to follow straightforwardly.

\subsection{$U(1)^2$ theory-II in general $D$ dimensions}
\label{sec:lu}

Another $U(1)^2$ gauged theory was proposed in \cite{Lu:2013eoa}.  The difference between this and the one in the previous subsection is that this theory involves only one scalar rather than two, and furthermore, the scalar can decouple and give rise to the Einstein-Maxwell theory. The charged AdS black hole is given by \cite{Lu:2013eoa}
\bea
ds^2 &=& - (H_1^{N_1} H_2^{N_2})^{-\fft{D-3}{D-2}} \, \tilde f\, dt^2 +
 (H_1^{N_1} H_2^{N_2})^{\fft{1}{D-2}}\big(\tilde f^{-1} dr^2 + r^2 d\Omega_{D-2}^2\big)\,,\nn\\
A_i &=& \sqrt{\ft{N_1(\mu+k q_i)}{q_i}} H_i^{-1} dt\,,\qquad
\phi= \ft12 N_1 a_1 \log H_1 + \ft12 N_2 a_2 \log H_2\,,\nn\\
\tilde f &=& k - \fft{\mu}{r^{D-3}} + g^2 r^2 H_1^{N_1} H_2^{N_2}\,,\qquad H_i = 1 + \fft{q_i}{r^{D-3}}\,,
\eea
where
\bea
a_i^2 &=& \fft{4}{N_i} - \fft{2(D-3)}{D-2}\,,\qquad a_1 a_2 = -\fft{2(D-3)}{D-2}\,,\nn\\
&&N_1 + N_2 = \fft{2(D-2)}{D-3}\,.
\eea
The solution with $q_2=0$ was also obtained in \cite{Gao:2004tu}.  The general solution has three integration constants $(\mu, q_1, q_2)$, parameterizing the mass and two electric charges \cite{Lu:2013eoa}
\bea
M &=& \fft{\Omega_{D-2}}{16\pi}\big( (D-2)\mu + (D-3)k (N_1 q_1 + N_2 q_2)\big),\nn\\
Q_i &=&  \fft{(D-3) \Omega_{D-2}}{16\pi} \sqrt{N_i q_i (\mu+k q_i)}\,.
\eea
The solution describes a black hole when there exists $r_0>0$ such that $\tilde f(r_0)=0$.
The thermodynamical variables are
\bea
T &=& \fft{\hat f'}{4\pi \sqrt{H_1^{N_1} H_2^{N_2}}}\Big|_{r=r_0}\,,\qquad S=\ft14 \Omega_{D-2} \, \rho^{D-2}\,,\nn\\
\Phi_i &=&\fft{\sqrt{N_i q_i (\mu + k q_i)}}{r_0^{D-3} + q_i}\,,\qquad
P_{\rm th}= \fft{(D-1)(D-2) g^2}{16\pi}\,,\nn\\
V_{\rm th} &=& \fft{(D-3) \Omega_{D-2}}{2(D-1)(D-2)} r_0^{D-1} H_1^{N_1-1} H_2^{N_2-1} (N_2 H_1 + N_1 H_2)\,.\label{dilvth}
\eea
These quantities satisfy the first law
\be
dM= T dS + \Phi_1 dQ_1 + \Phi_2 dQ_2 + V_{\rm th} dP_{\rm th}\,.
\ee

\bigskip
\noindent{\bf Case 1: $q_1q_2 \ne  0$}
\medskip

In this case, there are two horizons, $r_+\ge r_->0$.  We find that the action growth rate is
\bea
&&\dot I_{\rm WDW} = (\Phi_1 Q_1+ \Phi_2 Q_2)\Big|_{r_+}^{r_-}=2 P_{\rm th} \Delta V_{\rm th} + \fft{k}{4\pi}(D-2) \Omega_{D-2} \big(
r_+^{D-3} - r_-^{D-3}\big)\nn\\
&&= 2M -\fft{(D-3)\Omega_{D-2}}{16\pi}\\
&&\times\left(\mu \sum_{i=1}^2
N_i \Big(\fft{r_-^{D-3}}{r_-^{D-3} + q_i} + \fft{q_i}{r_+^{D-3} + q_i}\Big)
+k \sum_{i=1}^2 N_i \Big(\fft{q_i^2}{r_+^{D-3} + q_i} + \fft{q_i (2 r_-^{D-3} + q_i)}{
r_+^{D-3} + q_i}\Big)\right).\nn
\eea
Thus the algebraic inequalities (\ref{ineq1}) are all satisfied. The general proof for the differential inequalities are difficult analytically since it involves derivatives of $r_-$ with respect to $r_+$. The parameters $(N_1,N_2)$ in general are not integers beyond five dimensions and the horizon equations $\tilde  f(r_+)=0=\tilde  f(r_-)$ are hard to solve analytically.  However, we can establish the differential inequality when the black hole has only one horizon, which we shall discuss next.

\bigskip
\noindent{\bf Case 2: $q_2 =  0$}
\medskip

When $N_1>(D-1)/(D-3)$, the solutions have two horizons and the analysis and the conclusion is the same as in case 1.  For $N_1\le  (D-1)/(D-3)$, there is only one horizon and we find
\bea
N_1 = \fft{D-1}{D-3}:&&\qquad \dot I_{\rm WDW} = 2M - \Phi_1 Q_1 - \fft{(D-3)\Omega_{D-2}}{16\pi}\Big(
k N_1 q_1 + g^2 q_1^{\fft{D-1}{D-3}}\Big),\nn\\
N_1 < \fft{D-1}{D-3}:&&\qquad \dot I_{\rm WDW} = 2M- \Phi_1 Q_1-
\fft{(D-3)\Omega_{D-2}}{16\pi} k N_1 q_1\,,\label{n1dotIwdw}
\eea
Note also that we have
\bea
N_1 = \fft{D-1}{D-3}:&&\qquad V_{\rm th}^0= \frac{(D-3)\Omega_{D-2}}{2 (D-2) (D-1)} q_1^{\frac{D-1}{D-3}}>0\,,\nn\\
N_1 < \fft{D-1}{D-3}:&&\qquad V_{\rm th}^0=0\,.
\eea
The volume at the singularity has a discontinuity when we change $N_1$, which is not an integration constant but a parameter specifying the theory. It is easy to verify that
\be
\dot I_{\rm WDW} - 2 P_{\rm th}\, \Delta V_{\rm th} = \ft{1}{8\pi} (D-2) \Omega_{D-2}\,k\, r_+^{D-3}\,.
\ee
Thus all the relations in (\ref{ineq1}) are satisfied.

We now examine the differential bound (\ref{diffineq}).  For the case with $N_1=(D-1)/(D-3)$, we find that $C=2$.  To see this explicitly, we define $Z$ as
\be
Z\equiv \fft{1}{T}\fft{\partial \dot I_{\rm WDW}}{\partial S}\Big|_{Q_1,P_{\rm th}}- 2\,.
\ee
We find that for $k=0$, we have
\be
Z=\fft{q_1}{D-2} r^{1-D} \Big(
(D-1) q_1^{\fft{2}{D-3}} H_1^{\fft{1-D}{D-3}} + (D-3) H_1^{-1} \Big)\Big|_{r=r_+}\,.
\ee
The expression for $k=1$ is more complicated. Writing $Z=X(r_+)/Y(r_+)$, we find
\bea
X&=& (D-1) g^4 q_1 r^2 H_1^{\frac{2}{D-3}} \left((D-3) r^2 H_1^{\frac{2}{D-3}}+(D-1) q_1^{\frac{2}{D-3}}\right)\nn\\
&&+(D-1) g^2 q_1 \left((3 D-7) r^2 H_1^{\frac{2}{D-3}}+(D-3) q_1^{\frac{2}{D-3}}\right)+2 (D-3) (D-1) q_1\,,
\nn\\
Y&=& \frac{4 \pi H_1^{\frac{1}{D-3}+\frac{1}{2}} r^{D-2}}{D-1} T
\Big(4 \pi  (D-2) r T\, H_1^{\frac{1}{D-3}+\frac{3}{2}}+(D-3) H_1+D-1\Big),
\eea
It is clear that the quantity $Z$ is nonnegative for both $k=0,1$, confirming the bound (\ref{diffineq}) with $C=2$.

We now examine (\ref{diffineq2}) for $k=1$.  We find
\be
\fft{1}{T} \fft{\partial (2P_{\rm th}\, \Delta V_{\rm th})}{\partial S}\Big|_{Q_1,P_{\rm th}}=\fft{(D-1) g^2 r_+^2 X}{Y}>0\,,
\ee
where
\bea
X&=& g^2 r_+^2 H_1^{\frac{9}{D-3}+2} q_1^{\frac{1}{3-D}} \Big(H_1^{\frac{2}{D-3}} q_1^{\frac{1}{D-3}} \left((3 D-7) q_1 r_+^3+2 (D-2) r_+^D\right)\nn\\
&&\qquad\qquad+(D-1) r_+ q_1^{\frac{D}{D-3}}\Big)\nn\\
&&+ H_1^{\frac{2 D+1}{D-3}} \left(H_1^{\frac{2}{D-3}} \left(3 (D-3) q_1 r_+^3+2 (D-2) r_+^D\right)+(D-3) r_+ q_1^{\frac{D-1}{D-3}}\right),\cr
Y&=& H_1^{\frac{2 D}{D-3}} r_+^D \left((D-1) g^2 r_+^2 H_1^{\frac{2}{D-3}}+D-3\right) \nn\\
&&\times \left((D-2) g^2 r_+^2 H_1^{\frac{D}{D-3}}+(D-3) H_1^{\frac{1}{D-3}+1}+H_1^{\frac{1}{D-3}}\right).
\eea

For $N_1<(D-1)/(D-3)$, we have $V_{\rm th}^0=0$ and $\dot I_{\rm WDW}$ was given in (\ref{n1dotIwdw}).
It is advantageous to write
\be
N_1 = \fft{D-1}{D-3} \fft{x}{1+x}\,,
\ee
where $x\in [0,\infty)$. For $k=0$, we find
\bea
&&\fft{1}{T} \fft{\partial \dot I_{\rm WDW}}{\partial S}\Big|_{Q_1,P_{\rm th}} - \fft{D-3}{D-2}\nn\\
&&=\frac{2 q_1 (x+1) r_+^{D+3} ((D-2) x+D-1)+(D-1) q_1^2 r_+^6+(D-1) (x+1)^2 r_+^{2 D}}{(D-2) (x+1) \left(r_+^D+q_1 r_+^3\right) \left((x+1) r_+^D+q_1 r_+^3\right)}.
\eea
Thus the bound (\ref{diffineq}) is satisfied with $C=(D-3)/(D-2)$.  For $k=1$, the expressions are more complicated, but the lower bound is the same. We find
\be
\fft{1}{T} \fft{\partial \dot I_{\rm WDW}}{\partial S}\Big|_{Q_1,P_{\rm th}} - \fft{D-3}{D-2} =
\fft{(D-2)X}{(D-3)H_1 Y}\,,
\ee
where
\bea
X&=&(D-2)g^4 r_+^4 H_1^{2N_1}\Big( (D-1) \left(H_1^2+x (x+2)\right)\nn\\
&&\qquad +2 q_1 x r_+^{3-D} ((D-2) x+2 D-3)\Big)\nn\\
&&+g^2 r_+^2 H_1^{1 + N_1}
\Big(q_1 (x+1) r_+^{3-D} \left(\left(5 D^2-22 D+23\right) x+2 \left(2 D^2-7 D+6\right)\right)\nn\\
&&+(D-1) q_1^2 r_+^{6-2 D} ((D-3) x+2 (D-2))+2 (D-2)^2 (x+1)^2\Big)\nn\\
&&+(D-3) H_1^2 q_1 (x+1) r_+^{3-D} ((3 D-7) x+2 (D-2))\nn\\
&&+(D-3) (D-2) H_1^2 (x+1)^2\,,\nn\\
Y&=&\Big((D-1) g^2 r_+^2 H_1^{N_1} \left(H_1+x\right)+(D-3) H_1 (x+1)\Big)\nn\\
&&\times \Big((D-2) (x+1) \left(g^2 r_+^2 H_1^{N_1}+1\right)+q_1 r_+^{3-D} ((D-3) x+2 (D-2))\Big)\,.
\eea

We now consider the relation (\ref{diffineq2}) and for $k=1$ we find
\be
\fft{1}{T} \fft{\partial (2P_{\rm th}\, \Delta V_{\rm th})}{\partial S}\Big|_{Q_1,P_{\rm th}}=g^2 r_+^2 H_1^{N_1}\fft{ X}{Y}\,,
\ee
where
\bea
X&=&\fft{(D-1)g^2 r_+^2 H_1^{N_1}}{(1+x)^2}
\Big(H_1^2 ((D-3) x+2 (D-2))\nn\\
&&+q_1x r_+^{3-D} ((3 D-7) x+5 D-9)+x (2 (D-2) x+3 D-5)\Big)\nn\\
&& + \fft{(D-1)H_1}{(1+x)^2}\Big(2 (D-2) (x+1)^2\nn\\
 &&+q_1 ((D-3) x+2 (D-2)) \big(3 (x+1) r_+^{3-D}+2 q_1 r_+^{-2D}\big)\Big)\,,\nn\\
Y&=& \fft{H_1}{(1+x)^2}\Big((D-1) g^2 r_+^2 H_1^{N_1} \left(H_1+x\right)+(D-3) H_1 (x+1)\Big)\nn\\
&&\times \Big((D-2) g^2 r_+^2 (x+1) H_1^{N_1}+ q_1 r_+^{3-D} ((D-3) x+2 (D-2))\nn\\
&&\qquad+(D-2) (x+1)\Big)\,.
\eea
Thus we see that the inequality (\ref{diffineq2}) also holds in this case.

\section{Einstein-Born-Infeld theory}
\label{sec:EBI}

Black holes in Einstein-Born-Infeld (EBI) theory in four dimensions was constructed in \cite{GSP1984}.
The solution was generalized to general higher dimensions \cite{Dey:2004yt} and topologies \cite{Cai:2004eh}.
For simplicity, we shall focus only on the $D=4$ case to examine the inequalities.  We follow the notations in \cite{Li:2016nll}, but we turn off the magnetic charge.  The solution is
\bea
ds^2 &=& - f dt^2 + \fft{dr^2}{f} + r^2 d\Omega_{2,k}^2\,,\qquad A=a dt\,,\qquad a= \fft{q}{r} \ {}_2F_1[\ft14,\ft12; \ft54; - \ft{q^2}{b^2 r^4}]\,,\nn\\
f &=& -\ft13 \Lambda_0 r^2 + k - \fft{\mu}{r} - \ft16 b^2 \sqrt{r^2 + \fft{q^2}{b^2}} +
\fft{q^2}{3r^2}\ {}_2F_1[\ft14,\ft12; \ft54; - \ft{q^2}{b^2 r^4}]\,.
\eea
The solution has two integration constants $(\mu, q)$ parameterizing the mass and the electric charge
\be
M=\ft12 \mu\,,\qquad Q=\ft14 q\,.
\ee
For sufficiently large $M$, the solution admits a horizon $r_0$ such that $f(r_0)=0$.  The first law of thermodynamics is
\be
dM=TdS + \Phi dQ + V_0 dP_0 + V_b dP_b\,,
\ee
where
\bea
T &=& \fft{f'(r_0)}{4\pi}\,,\qquad S=\pi r_0^2\,,\qquad P_0 =-\ft{\Lambda_0}{8\pi}\,,\qquad
V_0=\ft43 \pi r_0^3\,,\nn\\
P_b &=& -\fft{b^2}{16\pi}\,,\qquad V_b = \ft43\pi r_0^3\left(
\sqrt{\ft{q^2}{b^2 r_0^4}+1}-\frac{q^2 }{2 b^2 r_0^4} \, _2F_1\left[\ft{1}{4},\ft{1}{2};\ft{5}{4};-\ft{q^2}{b^2 r_0^4}\right]\right).
\eea
The AdS radius $\ell=1/g$ is given by $\Lambda_0=(-3g^2 - \fft12 b^2)$.  We can write $b=\beta g$ and fix the dimensionless quantity $\beta$, we then have $V_0 d P_0 + V_b dP_b = V_{\rm th} d P_{\rm th}$ with $P_{\rm th}=
3g^2/(8\pi)$ and
\be
V_{\rm th} = (1 + \ft16 \beta^2) V_0 + \ft16 \beta^2 V_b\,.
\ee
Then thermodynamical quantities satisfy the same Smarr relations (\ref{smarr1}), (\ref{smarr2}) and (\ref{smarr3}).  Here $V_{\rm th}$ is formally a function of the horizon radius $r_0$. We can now promote $V_{\rm th}(r_0)$ to a general local expression of $V_{\rm th}(r)$ as the volume of any radius $r$. To be precise, we should redefine the parameters $q=\tilde q g$ and $b=\beta g$ so that the general volume formula
$V_{\rm th}(r)$ is independent of the pressure $P_{\rm th}=3g^2/(8\pi)$.  We can now obtain the volume at the singularity. We find that it does not vanish, but is
\be
V_{\rm th}^0= V_{\rm th}(0)= \frac{\sqrt{\pi b q^3} \Gamma \left(\frac{1}{4}\right) \Gamma \left(\frac{5}{4}\right)}{9 g^2}\,.\label{ebivol}
\ee
Thus the volume of singularity is positive.

The singularity structure of the charged black hole is much complicated than that of the RN black hole.  Black holes with two horizons or simply just one horizon can all emerge.  When the black hole have inner and outer horizons, the action growth rate is
\be
\dot I_{\rm WDW} = \Phi Q \Big|^{r_-}_{r_+} = 2 P_{\rm th}\, \Delta V_{\rm th} + k (r_+ - r_-)\,.
\ee
Other inequalities are difficult to prove analytically since we need to parameterize both inner and outer horizons. The situation becomes easier when there is only one horizon, for which the action growth rate was obtained in \cite{cai2}. We find that its relation with $\Delta V_{\rm th}$ is the same as before, namely
\bea
\dot I_{\rm WDW} &=& 2M- \Phi Q -\frac{\sqrt{bq^3}\, \Gamma \left(\frac{1}{4}\right) \Gamma \left(\frac{5}{4}\right)}{12 \sqrt{\pi }}\nn\\
&=& 2P_{\rm th} \Delta V_{\rm th} + k r_+\,,
\eea
where $\Delta V_{\rm th}= V_{\rm th}(r_+) - V_{\rm th}^0$, where $V_{\rm th}^0$ is given by (\ref{ebivol}).  Thus the relations in (\ref{ineq1}) are all satisfied.

We can also prove analytically the differential bound (\ref{diffineq}) for solutions with single horizon.  We present the proof in detail for $k=0$. In this case, we have special values of mass parameter
\be
\mu_2= \frac{\sqrt{b q^3}\, \Gamma \left(\frac{1}{4}\right) \Gamma \left(\frac{5}{4}\right)}{3 \sqrt{\pi }} \ge \mu_1=
\frac{\sqrt{2gq^3}\, \sqrt[4]{b^2+3 g^2} }{3^{3/4} \sqrt{b}}
\, _2F_1\left[\ft{1}{4},\ft{1}{2};\ft{5}{4};-\ft{12 g^2 \left(b^2+3 g^2\right)}{b^4}\right].
\ee
The inequality is saturated at $b=0$.  We find
\bea
\mu>\mu_2:&&\qquad \hbox{one horizon}\,,\nn\\
\mu_1<\mu<\mu_2:&&\qquad \hbox{two horizons}\,,\nn\\
\mu=\mu_1:&&\qquad \hbox{extremal, with\ } r_\pm=\frac{\sqrt{bq}}{\sqrt[4]{3} \sqrt{2g} \sqrt[4]{b^2+3 g^2}}\,,\nn\\
\mu<\mu_1:&&\qquad\hbox{naked singularity.}
\eea
We now focus on the $\mu\ge \mu_2$ case and keep the charge $Q, P_{\rm th}$ and $\beta$ fixed, we find
\bea
\fft{1}{T} \fft{\partial \dot I_{\rm WDW}}{\partial S}\Big|_{Q,P_{\rm th}} = 2+
\frac{b q^2}{8 \pi  r_+ T \sqrt{b^2 r_+^4+q^2}}\ge 2\,.
\eea
Exact same result can be achieved for $k=1$ case when there is only one horizon. (In fact the form of the result is independent of $k$.) We thus prove the bound (\ref{diffineq}) with $C=2$ when the charged black hole has only one horizon.

     For $k=1$, for black holes with a single horizon, we find
\bea
\fft{1}{T} \fft{\partial (2P_{\rm th}\, \Delta V_{\rm th})}{\partial S}\Big|_{Q,P_{\rm th}}=\fft{1}{8\pi r_+\, T}\Big(
2 \left(\beta ^2+6\right) g^2 r_+^2-\frac{\beta g  \left(2 \beta ^2 g^2 r_+^4+q^2\right)}{\sqrt{\beta ^2 g^2 r_+^4+q^2}}\Big)\,.
\eea
This quantity however can be negative for sufficiently small $r_+$. To be precise, the quantity becomes negative when the horizon radius lies in the range
\be
0< r_+< \frac{\sqrt{\beta q}}{\sqrt{2g} \sqrt[4]{6 \left(\beta ^2+3\right)+\sqrt{3} \left(\beta ^2+6\right) \sqrt{\beta ^2+3}} }\,.
\ee
This is the second example where the inequality (\ref{diffineq2}) breaks down for $k=1$.

When the black holes have two horizons, the differential relation (\ref{diffineq}) is much more difficult to demonstrate.  However, the partial derivative of $\dot I_{\rm WDW}$ with respect to the entropy can be obtained straightforwardly, given by
\bea
&&\fft{1}{T}\fft{\partial \dot I_{\rm WDW}}{\partial S}\Big|_{Q,P_{\rm th}} =
\frac{b q^2}{64 \pi ^2 r_- r_+ (-T_-) T_+\sqrt{b^2 r_-^4+q^2} \sqrt{b^2 r_+^4+q^2}}\Big[\nn\\
&&6g^2 \left( r_+^2 \sqrt{b^2 r_+^4+q^2}- r_-^2 \sqrt{b^2 r_-^4+q^2}\right)+k \left(2 \sqrt{b^2 r_+^4+q^2}-2 \sqrt{b^2 r_-^4+q^2}\right)\nn\\
&&+b^2 \left(r_+^2 \sqrt{b^2 r_+^4+q^2}-r_-^2 \sqrt{b^2 r_-^4+q^2}\right)-b^3 \left(r_+^4-r_-^4\right)\Big]\,.\label{ebires}
\eea
Note that $T_-<0$ and $T_+>0$ denote the temperature in inner and outer horizons respectively and $T$ in the left-hand side of the equation is understood to be always $T_+$.  Owing to the fact that the horizons are determined by $f(r_0)=0$ which involves a hypergeometric function, we do not have an analytical method to validate the relation (\ref{diffineq}).  By going through the parameter space numerically, we find that the relation (\ref{diffineq}) is indeed satisfied with $C=3/2$.  Our numerical approach is the following. We first solve for $(g^2,\mu)$ in terms of $(r_+,r_-,b,q)$. We find that for given $(r_+,r_-,b)$, we must have $q>q_{\rm min}$ for some $q_{\rm min}(r_+,r_-,b)$ so that $(g^2, \mu)$ are both positive.  We find that (\ref{ebires}) is a monotonically decreasing function of $q>q_{\rm min}$.  The quantity at $q\rightarrow \infty$ depends on $r_+/r_-$ only and achieves the minimum $3/2$ when $r_+/r_-\rightarrow \infty$.

\section{Negative volume and violation of Lloyd's bound}
\label{sec:negvol}

In the section \ref{sec:bounds}, we introduced the concept of volume of spacetime singularity.  The concept of the area of the horizon is a special case of the more general areas of hyper surfaces of different radii.  We generalize the thermodynamic volume $V_{\rm th}(r_+)$ to a quantity $V_{\rm th}(r)$ that is a function of the generic radial coordinate $r$.  This generalization was done first in \cite{Feng:2017wvc}. It is perhaps misnomer to call these quantities ``area'' or ``volume'' inside the WDW patch where $r$ becomes a timelike coordinate; however, we shall continue to use the terminologies for lacking of better descriptions.

For many black holes, including Schwarzschild-AdS and RN-AdS with $V_{\rm th}(r) =\fft43\pi r^3$ in four dimensions, the volume vanishes at the spacetime singularity $r=0$.  However the quantity $V_{\rm th}^0$ at the singularity does not have to vanish and we found concrete examples in both EMD and EBI theories in the previous sections.  Furthermore we found that the action growth rate for the $k=0$ black holes with single horizon could be in general expressed as
\be
\dot I_{\rm WDW} = 2M - \Phi_i Q_i - 2P_{\rm th} V_{\rm th}^0\,.
\ee
Since the quantities $\Phi_i Q_i$ and $P_{\rm th}$ are positive, the Lloyd's bound is satisfied provided that $V_{\rm th}^0\ge 0$.  The violation of Lloyd's bound is then possible for black holes with $V_{\rm th}^0<0$.
In both the EMD and EBI theories we considered in the previous two sections, we had $V_{\rm th}^0\ge 0$ and hence the Lloyd's bound were all satisfied.

In this section, we provide some concrete examples in Einstein-scalar theories that give rise to black holes with negative $V_{\rm th}^0$.  The Lagrangian takes the general form
\be
{\cal L}=\sqrt{-g} \Big(R - \ft12 (\partial \phi)^2 - V(\phi)\Big)\,,
\ee
where $V(\phi)$ is the scalar potential.  We consider two scalar potentials. One is given by \cite{Fan:2015tua}
\bea
V &=& -\ft12 g^2 (D-2)\big(D + (D-2) \cosh(\lambda\phi)\big) \nn\\
&&- \ft12 \alpha (D-2) (e^{\lambda\phi}-1)^{D-1}\Big((D-1)
(1 + e^{-\lambda\phi})e^{-\fft{\phi}{\lambda}}\nn\\
&&\qquad \qquad-\big(D+(D-2)\cosh(\lambda\phi)\big)\,
{}_2F_1[D-1,\ft12D; D;1-e^{\lambda\phi}]\Big)\,,\label{scalarpot1}
\eea
where $\lambda=\sqrt{2/(D-2)}$. Note that the hypergeometric function reduces to simpler functions for integer $D$. Note that the parameter $g^2$ in the potential should not be confused with the determinant of the metric. The other is \cite{Feng:2013tza}
\bea
V(\phi) &=& - \ft12 (D-2) g^2 e^{\fft{\mu-1}{\nu}\Phi}\Big[
(\mu-1)((D-2)\mu-1) e^{\fft{2}{\nu}\Phi} -2(D-2)(\mu^2-1) e^{\fft{1}{\nu}\Phi}\cr
&&\qquad\qquad\qquad\qquad\quad + (\mu+1)((D-2)\mu+1)\Big]\cr
&&-\ft{(D-3)^2}{2(3D-7)}(\mu+1) \alpha\, e^{-\fft{1}{\nu} (4+\fft{\mu+1}{D-3})\Phi} (e^{\fft{1}{\nu}\Phi} -1)^{3+\fft{2}{D-3}}\cr
&&\times\Big[(3D-7) e^{\fft{1}{\nu}\Phi}\,{}_2F_1[2,1+\ft{(D-2)(\mu+1)}{D-3};
3+\ft{2}{D-2};1 - e^{\fft1{\nu}\Phi}] \nn\\
&& - \big((3D-7) + (D-2)(\mu-1)\big)\,
{}_2F_1[3,2+\ft{(D-2)(\mu+1)}{D-3};4+\ft{2}{D-2};1 - e^{\fft1{\nu}\Phi}]
\Big],\label{scalarpot2}
\eea
where $\mu^2+\nu^2=1$ and
\be
\Phi=\sqrt{\ft{2(D-3)}{D-2}}\,\phi\,.\label{Phidef}
\ee
In $D=4,5$, the hypergeometric function reduces to simpler functions and the $D=4$ theory was first obtained in \cite{Anabalon:2013qua}. In both of the above scalar potentials, there is a fixed point at $\phi=0$ with
\be
V(0)=-\ft12 (D-1)(D-2) g^2\,,
\ee
giving rise to AdS vacua of radius $\ell=1/g$.  The Taylor expansion of the scalar around $\phi=0$ indicates that $\phi$ is conformally massless.  The parameter $\alpha>0$ has the same dimension of $g^2$ and the $\alpha$-term is necessary for the construction of the black hole solution.  The complete first law of black hole thermodynamics then necessary include $\alpha$, namely
\be
dM=TdS + V_g dP_{\rm th} + V_\alpha d\alpha\,.
\ee
In this paper, we shall unify the description so that the Smarr relations discussed in section \ref{sec:bounds} would hold. We let $\alpha=\beta g^2>0$, where the dimensionless $\beta$ is fixed thermodynamically. We define
\be
V_{\rm th} = V_g + \fft{16\pi\beta}{(D-1)(D-2)} V_{\alpha}\,.
\ee
The first law then becomes
\be
dM=TdS + V_{\rm th} dP_{\rm th}\,.
\ee
Since the parameter $\beta$ is dimensionless, it follows that it does not involve in both the Smarr and generalized Smarr relations discussed in section \ref{sec:bounds}.  It is also important to require that when we promote $V_{\rm th}(r_+)$ to general $V_{\rm th}(r)$, the local volume expression $V_{\rm th}(r)$ is independent of the pressure $P_{\rm th}$.

\subsection{Example 1}

In this subsection, we consider the scalar potential (\ref{scalarpot1}).  For simplicity, we begin with the illustration in $D=4$.  The potential is \cite{Zloshchastiev:2004ny}
\be
V(\phi)=- 2 g^2 \Big((\cosh\phi+ 2)- 2\beta^2(2\phi + \phi \cosh\phi - 3\sinh\phi)\Big)\,.
\ee
The parameter $\beta$ is a fixed dimensionless quantity.  The theory admits asymptotic AdS black hole, given by \cite{Zloshchastiev:2004ny}
\bea
ds^2 &=& -f dt^2 + \fft{dr^2}{f} + r(r+q) d\Omega_{2,k}^2\,,\qquad e^\phi=1 + \fft{q}{r}\,,\nn\\
f &=& g^2r^2 + k - \ft12g^2 \beta^2 q^2 + g^2(1-\beta^2) q r + g^2 \beta^2 r^2 \big(1 + \fft{q}{r}\big) \log\big(1 + \fft{q}{r}\big)\,.
\eea
(The static solution can be prompted to be time dependent, describing exact formation of the black hole \cite{Zhang:2014sta}.) The solution contains only one integration constant $q$, parameterizing the mass
\be
M=\ft1{12}g^2\beta^2 q^3\,.
\ee
For sufficiently large $M$ so that
\be
\ft12 g^2 \beta^2 q^2 -k >0\,,\label{case1bhcons}
\ee
the solution describes a black hole with single horizon $r_+>0$ satisfying $f(r_+)=0$ and the curvature singularity is located at $r=0$. For the planar boundary ($k=0$), the solution forms a black hole as long as $M>0$.  It can be easily verified that the first law $dM=TdS + V_{\rm th} dP_{\rm th}$ is satisfied, with
\bea
T &=& \fft{f'(r_+)}{4\pi}\,,\qquad S=\pi r_+(r_++q)\,,\qquad
P_{\rm th}=\fft{3g^2}{8\pi}\,,\nn\\
V_{\rm th} &=& \frac{2}{3} \pi  r_+^3 \big(1 + \fft{q}{r_+}\big) \big(2 + \fft{q}{r_+}\big) \left(1+\beta ^2 \log \big(1+\frac{q}{r_+}\big)\right)\nn\\
&&-\frac{1}{9} \pi  \beta ^2 q \left(q^2+12 q r_++12 r_+^2\right)\,.
\eea
Thus we see that the singularity volume is negative, namely
\be
V_{\rm th}^0=-\ft19 \pi \beta^2 q^3\,.
\ee
It should be pointed that since the parameters satisfying $f(r_+)=0$, the expression of $V_{\rm th}(r_+)$ is not unique. When we promote $V_{\rm th}(r_+)$ to $V_{\rm th}(r)$, different expressions of $V_{\rm th}(r_+)$ lead to different $V_{\rm th}(r)$ and hence different $V_{\rm th}^0$. We resolve this ambiguity by requiring that $V_{\rm th}(r)$ be independent of $g^2$ or the pressure.  An added benefit is that the expression $V_{\rm th}(r)$ remains unchanged even when we turn off the effective cosmological constant, making the earlier analogy of $V_{\rm th}(r)$ to the area of constant radius more appropriate.

Having negative volume at $r=0$ implies that for sufficiently small horizon radius $r_+$, the thermodynamical volume will become negative also. The solution however satisfies the null energy condition both inside and outside the horizon.  To see this explicitly, we define $K_0=g^{tt} G_{tt}$, $K_1=g^{rr} G_{rr}$ and $K_2=K_3=g^{22} G_{22}$ where $G_{\mu\nu}$ is the Einstein tensor.  We find
\bea
\hbox{outside the horizon:}&& -K_0 + K_1 = \fft{q^2 f}{2r^2(r+q)^2}\,,\qquad -K_0 + K_2=0\,.\nn\\
\hbox{inside the horizon:}&&
-K_1 + K_0= \fft{q^2 (-f)}{2r^2(r+q)^2} =-K_1 + K_2\,.
\eea
Thus the black hole satisfies the null energy condition for all parameters, demonstrating that the energy condition allows black holes to have negative thermodynamical volumes.

It turns out that the action growth rate in the WDW patch is
\be
\dot I_{\rm WDW} = 3M - \ft12 k q = 2M - \ft12 k q - 2 P_{\rm th} V_{\rm th}^0\,,
\ee
where the first equality above was given in \cite{cav6}. Thus the Lloyd's bound is clearly violated and the violation is closely related to the fact that the black hole has negative singularity volume.  The inequalities associated with $2P_{\rm th}\, \Delta V_{\rm th}$ however remain, since we have
\be
\dot I_{\rm WDW} - 2P_{\rm th}\, \Delta V_{\rm th} =k r_+\,.
\ee

Even though the black hole violates the Lloyd's bound, we find that the differential bound is still satisfied. Specifically, we find
\be
\fft{1}{T}\fft{\partial \dot I_{\rm WDW}}{\partial S}\Big|_{P_{\rm th}}=
3 - \fft{2k}{\beta^2 g^2 q^2}\qquad
\left\{
  \begin{array}{ll}
    \ge 2, &\qquad k=1\,, \\
    =3, &\qquad k=0\,.
  \end{array}
\right.
\ee
The inequality associated with $k=1$ is a consequence of the black hole constraint (\ref{case1bhcons}).

We also find
\be
\fft{\partial (2P_{\rm th}\, \Delta V_{\rm th})}{\partial S}\Big|_{P_{\rm th}}
=\fft{1}{8\beta^2 g^2 \pi q r_+ (r_+ + q)}\Big(
(\beta^2 g^2 q^2 - 2) (3\beta^2 g^2 q^2-2) -16 \beta^2 g^2 q r_+\Big)\,.
\ee
This quantity is not manifestly positive; however, we have checked large number of numerical quantities and find that it is always positive.

When $k=0$, this theory and the solution can be generalized to general dimensions and the scalar potential was given in (\ref{scalarpot1}). The metric of the solution is given by \cite{Fan:2015tua}
\bea
ds^2 &=&  - f dt^2 + \fft{dr^2}{f} + r(r+q) dx^i dx^i\,,\nn\\
f &=& r(r+q) \Big(g^2 - \fft{\alpha q^{D-1}}{r^{D-1}}\, {}_2F_1[D-1, \ft12 D; D; -
\ft{q}{r}]\Big)\,.
\eea
The mass of the black hole is
\be
M=\fft{\Omega_{D-2}}{16\pi} (D-2) \beta g^2\,.
\ee
The black hole has only one horizon $r_+$ and we find that the volume formula is
\bea
V_{\rm th} &=& \frac{\Omega _{D-2} \left(q+2 r_+\right) \left(r_+ \left(q+r_+\right)\right){}^{\frac{D}{2}-1}}{2 (D-1)} + \frac{\Omega _{D-2} \beta  q^{D-1} r_+^{-\frac{D}{2}}}{2 (D-1) \left(q+r_+\right)}\Big[\nn\\
&&2 r_+^{\fft{D}2} \left(q+r_+\right)-\left(q+2 r_+\right) \left(q+r_+\right)^{\fft{D}2} \, _2F_1[D-1,\ft12D;D;-\ft{q}{r_+}]\Big]\,.
\eea
Note that we have negative volume at singularity, given by
\be
V_{\rm th}^0=-\frac{\beta  q^{D-1} \Omega _{D-2}}{(D-2) (D-1)}\,.
\ee
We find that the action growth rate now becomes
\be
\dot I_{\rm WDW} = 2M - 2P_{\rm th} V_{\rm th}^0 = 2 P_{\rm th}\,
\Delta V_{\rm th}= \fft{2(D-1)}{D-2} M>2M\,.
\ee
Thus we see that owing to the negative volume at singularity, the Lloyd's bound is violated in this class of black holes, but our proposal of relation between the CA and the new CV conjectures remains valid. It is also clear that in this ($k=0$) case, we have
\be
\fft{1}{T} \fft{\partial \dot I_{\rm WDW}}{\partial S}\Big|_{P_{\rm th}} =
\fft{1}{T} \fft{\partial (2P_{\rm th}\,\Delta V_{\rm th})}{\partial S}\Big|_{P_{\rm th}}=
\fft{2(D-1)}{D-2}\,.
\ee

\subsection{Example 2}

We now consider scalar potential (\ref{scalarpot2}).  The Einstein-scalar theory admits neutral AdS black holes and some special exact solutions were found \cite{Feng:2013tza}. The metric of the solution is given by
\bea
ds^2 &=& - \fft{\tilde f}{H^{1+\mu}} dt^2 + H^{\fft{1+\mu}{D-3}} \Big(\fft{dr^2}{\tilde f} +
r^2 d\Omega_{D-2}^2\Big)\,,\qquad H=1 + \fft{q}{r^{D-3}}\,,\nn\\
\tilde f &=& g^2 r^2 H^{\frac{(D-2) (\mu +1)}{D-3}}+ k H\nn\\
&&-\beta g^2 r^2 (H-1)^{\frac{D-1}{D-3}} \, _2F_1\left(1,\frac{(D-2) (\mu +1)}{D-3};\frac{2 (D-2)}{D-3};1-\frac{1}{H}\right)\,.
\eea
The solution contains one integration constant $q$, parameterizing the mass of the solution, given by
\be
M=\frac{(D-2)\Omega_{D-2}\,q \left(\beta g^2 q^{\frac{2}{D-3}}+k \mu \right)}{16 \pi }\,.
\ee
Note that in many black holes discussed in the previous sections, we used $\mu$ as the mass parameter.  It is not the case here; $\mu$ is instead a parameter of the theory, introduced in \cite{Feng:2013tza}.  The condition for the solution to describe a black hole is
\bea
-1\le \mu < \fft{1}{D-2}:&&\qquad c\equiv -(k + \frac{\beta  (D-1) g^2 q^{\frac{2}{D-3}}}{(D-2) \mu -1})>0\nn\\
\fft1{D-2}\le \mu\le 1: &&\qquad \hbox{no further conditions}.\label{cdef}
\eea
These conditions ensure that the mass of all the black holes is positive. Note that the solutions have curvature singularity at $r=0$, except for $\mu=1$.  When $\mu=1$, the solution reduces to the Schwarzschild black hole and the curvature singularity is at $H=0$ rather than $r=0$.  We shall hence ignore the $\mu=1$ case here. The thermodynamical variables are
\bea
T &=& \frac{ H_+^{-\frac{(D-2) (\mu +1)}{2 (D-3)}}}{4 \pi }\tilde f_+',\quad
S=\fft{\Omega_{D-2}}{4} r_+^{D-2} H_+^{\frac{(D-2) (\mu +1)}{2 (D-3)}},\quad P_{\rm th} = \fft{(D-1)(D-2)g^2}{16\pi}\,,\nn\\
V_{\rm th} &=& \frac{\Omega_{D-2}\, r^{D-1} ((1-\mu )H+\mu +1) H^{\frac{(D-2) \mu +1}{D-3}}}{2 (D-1)}
+\fft{\Omega_{D-2}\, \beta \,q^{\fft{D-1}{D-3}} }{2(D-1) H}\Big(2 H\nn\\
&& - \big((1-\mu) (H-1) +2\big)\, {}_2F_1 [1, \ft{(1+\mu)(D-2)}{D-3}; \ft{2(D-2)}{D-3}; 1-H^{-1}]\Big)\,.
\eea
We now study the volume of singularity $r=0$.  For $-1\le \mu < 1/(D-2)$, we find that the volume is negative, given by
\be
V_{\rm th}^0=-\frac{\beta  (D-3) (\mu +1) q^{\frac{2}{D-3}+1} \Omega _{D-2}}{2 (D-1) (1-(D-2) \mu )}<0\,.
\ee
However, for $1/(D-2) <\mu< 1$, the volume of singularity diverges as $1/\sqrt{r}$, namely $V_{\rm th}^0$
\be
V_{\rm th}^0 =\frac{(1-\mu) \Omega _{D-2} q^{\frac{(D-2) (\mu +1)}{D-3}}}{2 (D-1) r^{(D-2) \mu-1}} +
\frac{\beta  (D-3) (\mu +1) q^{\frac{2}{D-3}+1} \Omega _{D-2}}{2 (D-1) ((D-2) \mu -1)},\qquad r\rightarrow 0\,.
\ee
Note that in the above analysis, we made use of the hypergeometric identity identity
\be
{}_2F_1 [1, \ft{(1+\mu)(D-2)}{D-3}; \ft{2(D-2)}{D-3}; 1]=\fft{D-1}{1-(D-2)\mu}\,.
\ee
It is now straightforward to verify the following identity of the black hole:
\be
2M =2 P_{\rm th} V_{\rm th} + \fft{\Omega_{D-2}}{16\pi} (D-2) k \Big(2 r_+^{D-3}+(\mu +1) q\Big)\,.
\label{example2id}
\ee
We evaluate the action growth rate of the WDW patch and find
\bea
\dot I_{\rm WDW} &=& 2M - 2P_{\rm th} V_{\rm th}^0 - \frac{(D-2) k (\mu +1) q \Omega _{D-2}}{16 \pi }\nn\\
&=& 2 P_{\rm th}\, \Delta V_{\rm th} + \frac{(D-2) \Omega _{D-2}}{8 \pi } k r_+^{D-3}\,.
\eea
The first equality came from the direct evaluation of the action and the second equality then follows from the identity (\ref{example2id}).  It should be pointed out that for $1/(D-2) <\mu< 1$, both $I_{\rm WDW}$ and $V_{\rm th}^0$ diverge, but they diverge in the exact same way at the singularity $r=0$ and hence the above equality holds.  Thus we find the concrete examples of black holes whose action growth rate in the WDW patch diverges owing to the divergence of the volume of singularity.

For $-1\le \mu < 1/(D-2)$, the volume of singularity $V_{\rm th}^0$ is finite and negative. It follows that Lloyd's bound is violated, but our proposal of the relations associated with $\dot I_{\rm WDW}$ and $2P_{\rm th}\, \Delta V_{\rm th}$ are still valid.

We now examine the differential inequality (\ref{diffineq}). This inequality only makes sense when $\dot I_{\rm WDW}$ is finite, for which we must have $-1\le \mu < 1/(D-2)$, (or $\mu=1$, corresponding to the Schwarzschild black hole.)  In this parameter region, it is advantageous to define
\be
\mu=\fft{1-(D-2)x}{(D-2)(1+x)}\,,\qquad \hbox{with}\qquad x\in (0, \infty]\,.
\ee
We find that
\be
\fft{1}{T} \fft{\partial \dot I_{\rm WDW}}{\partial S}\Big|_{P_{\rm th}} =2 +
\frac{c (D-3) (D-1)}{c (D-2) (D-1) x+k (2 (D-2) x+D-3)}\,.
\ee
Thus the bound (\ref{diffineq}) holds with $C=2$.  Note that the quantity $(c>0)$, defined in (\ref{cdef}), is associated with the integration constant of the solution, but $(x>0)$ is a reparametrization of the constant $\mu$ in the theory. Thus the lowest $C$ increases for a given $\mu$ when $k=0$. We also find that
\be
\fft{1}{T} \fft{\partial (2 P_{\rm th}\, \Delta V_{\rm th})}{\partial S}\Big|_{P_{\rm th}} =2+
\frac{(D-3) (D-1) (c+k)}{c (D-2) (D-1) x+k (2 (D-2) x+D-3)}.
\ee

\section{Conclusions}
\label{sec:con}

In this paper, we studied the holographic complexity based on the CA conjecture and evaluated the action growth rate $\dot I_{\rm WDW}$ of large classes of static AdS black holes, in Einstein gravities with minimally coupled matter that satisfies the null energy condition.  We also made the new CV conjecture where the later time complexity growth rate is instead $\dot {\cal C}_V=2P_{\rm th}\, \Delta V_{\rm th}$. For black holes with two horizons, $V_{\rm th}=V_{\rm th}^+ - V_{\rm th}^-$, {\it i.e.}~the difference of the thermodynamical volumes of outer and inner horizons.  For black holes with only single horizon, we introduced a concept of volume of spacetime singularity $V_{\rm th}^0$ and $\Delta V_{\rm th} = V_{\rm th}^+ - V_{\rm th}^0$.  We demonstrated that for general AdS black holes with single horizon, there was the following relation
\bea
\dot I_{\rm WDW} &=& 2M -\Phi_i Q_i - 2P_{\rm th} V_{\rm th}^0 +k u(r_+)\nn\\
&=&2 P_{\rm th}\, \Delta V_{\rm th} + k u(r_+)\,,
\eea
where $u(r_+)$ is a simple positive function of $r_+$ that vanishes when $r_+\rightarrow 0$. For the planar AdS black holes $(k=0)$, we simply have $\dot I_{\rm WDW}=2P_{\rm th}\, \Delta V_{\rm th}$. In other words, we have $\dot {\cal C}_A=\dot {\cal C}_V$. Thus we see that the CA and the new CV conjectures satisfy the relations (\ref{ineq1}), except for the Lloyd's bound that could be violated.

Our relation between the action growth rate and the volume of black hole singularity allows us to give a simple criterium for the Lloyd's bound.  The bound is satisfied provided that $V_{\rm th}^0$ is positive or zero, but it can be violated when $V_{\rm th}^0$ becomes negative. Black holes with negative $V_{\rm th}^0$ do exist and they typically arise in Einstein-scalar theories.  We provided two classes of such black holes in general dimensions. However, regardless whether the Lloyd's bound does or does not hold, the relations between $\dot I_{\rm WDW}$ and $2P_{\rm th}\, \Delta V_{\rm th}$ in (\ref{ineq1}) are always satisfied and we found no exception.

What is disturbing is that we found explicit AdS black holes whose on-shell action growth rate in the WDW patch were divergent, and the cause of the divergence could be related to the divergence of the volume of singularity in these black holes. The CA conjecture cannot apply in these cases and the phenomenon calls for further investigation.

It should be pointed out that the area of horizon, which can be calculated locally on horizon, is a special case of the more general area of any given radius. The area is a pure geometric concept and can be calculated from the metric without having to know the detail of the full theory. The generalization of the thermodynamical volume to a volume at any radius $r$, including the singularity, is not always obvious. For the Schwarzschild or the RN black holes, the volume is simply $V_{\rm th}=\ft43 \pi r_+^3$ in four dimensions and the generalization to $V_{\rm th}(r)=\ft43 \pi r^3$ at any $r$ requires no deep thoughts and it allows us to deduce that $V_{\rm th}^0=0$. For large classes of black holes such empirical local expression for the volume was obtained in \cite{Feng:2017wvc}. However, general valid local formula is still lacking.  It is then not always obvious how to promote the $V_{\rm th}^+$ that is valid only on the horizon to general $V_{\rm th}(r)$ since $r_+$ can be expressed in terms of other parameters of the solutions, through the horizon constraint $g_{tt}(r_+)=0$.  However, in all our examples considered in this paper, we found that we could resolve the ambiguity by requiring that $V_{\rm th}(r_+)$ and hence $V_{\rm th}(r)$ are independent of $g^2$ or the thermodynamical pressure.  The resulting $V_{\rm th}(r)$ then is applicable even for asymptotically flat black holes where the concept of thermodynamical volume no longer exists. These results are suggestive of some independent geometric calculation for the black hole volumes without using black hole thermodynamics, even for those that are asymptotic to the Minkowski spacetime.

Perhaps what is more intriguing is the ${\cal C}$-$S$ relation we found that emerged as a differential inequality (\ref{diffineq}).  All the $D\ge 4$ black hole examples in this paper satisfy the lower bound $C=(D-3)/(D-2)$.  We could analytically prove the inequality for the majority of the black holes discussed in this paper; for those we could not, we tried to verify with many numerical data and we found no exception.  The robustness indicates a deep relation between the complexity and entropy of a quantum system that was not known in quantum information theories. On the other hand, if we treat $2P_{\rm th}\, \Delta V_{\rm th}$ as the holographic complexity, we found that the bound was at best replaced by the inequality (\ref{diffineq2}). However, even this inequality can be violated. One example is provided by the two-charged black holes in $D=7$ gauged supergravity and another is the charged AdS black hole in Einstein-Born-Infeld theory. Establishing the analogous ${\cal C}$-$S$ relation in quantum information theories can validate the holographic complexity and distinguish the two conjectures.

    It is worth mentioning that we have only considered neutral or purely electrically-charged static black holes. To limit the scope of this paper, we have avoided the discussion of magnetic, dyonic and rotating black holes. There are additional subtleties including the issue of electromagnetic duality of the on-shell actions in these cases \cite{susk,goto,Liu:2019smx} and we would like to address them in a separate study.

     Finally we would like to emphasize that we have found the following relations
\be
\fft{1}{T} \fft{\partial \dot I_{\rm WDW}}{\partial S}\Big|_{Q, P_{\rm th}}> C\,,\qquad
\hbox{and}\qquad
\left\{
  \begin{array}{ll}
    \dot I_{\rm WDW}>2P_{\rm th}\, \Delta V_{\rm th}, &\qquad k=1, \\
    \dot I_{\rm WDW}=2P_{\rm th}\, \Delta V_{\rm th}, &\qquad k=0, \\
    \dot I_{\rm WDW}<2P_{\rm th}\, \Delta V_{\rm th}, &\qquad k=-1.
  \end{array}
\right.
\ee
These relations withstood the test by all of our large number of examples, with no exception. In particular the CA and the new CV conjectures are identically the same for planar AdS black holes. Thus, regardless the validity of the CA or CV conjectures, these robust relations of the black hole thermodynamical variables may indicate some universal deep structures of AdS black holes.

\section*{Acknowledgement}

H.S.-Liu is supported in part by NSFC (National Natural Science Foundation of China) Grant No.~11475148 and No.~11675144,  H.L., L.M.~and W.D.~Tan are supported in part by NSFC Grants
No.~11875200 and No.~11935009.

\end{document}